\documentclass[aps,prd,showpacs]{revtex4}%
\usepackage{amsmath}
\usepackage{graphicx}
\usepackage{epsfig}
\usepackage{amsfonts}
\usepackage{amssymb}%
\providecommand{\U}[1]{\protect\rule{.1in}{.1in}}
\begin{document}
\title{\ A proposal of a local modified QCD }
\author{A. Cabo Montes de Oca $^{1,2}$ \bigskip}

\address{$^{1}$ Programa de P\'os-Graduac\~ao em F\'isica (PPGF) da
Universidade Federal do Par\'a (UFPA), Av. Augusto Correa,
No.  01, Campus B\'asico do Guam\'a,  Bel\'em, Par\'a, Brasil
\bigskip}

\address{$^{2}$ Departamento de F\'{\i}sica Te\'orica,
Instituto de Cibern\'{e}tica, Matem\'{a}tica y F\'{\i}sica, Calle E,
No. 309, Vedado, La Habana, Cuba. }

\begin{abstract}
\noindent A local and renormalizable version of a modified PQCD introduced in previous works is presented. The construction indicates that it could be equivalent to massless QCD. The case in which only quark condensate effects are retained is discussed in more detail. Then, the appearing auxiliary fermion fields can be integrated leading to a theory with the action of massless QCD, to which  one local and gauge invariant Lagrangian term for each quark flavour is added. These terms are defined by two gluon and two quark fields, in a form curiously not harming power counting renormalizability. The gluon self-energy is evaluated in second order in the gauge coupling and all orders in the new quark couplings, and the result became transversal as required by the gauge invariance. The vacuum energy was calculated in the two loop approximation and also became gauge parameter independent. The possibilities that higher loop contributions to the vacuum energy allow the generation of a quark mass hierarchy as a flavour symmetry breaking effect are discussed. However, the decision on this issue needs the evaluation of more than two loop contributions, in which more than one type of quark loops start appearing, possibly leading to interference effects in the vacuum energy.

\end{abstract}

\maketitle

\section{Introduction}

The understanding of the hierarchy of quark masses is a fundamental open
problem of High Energy Physics. In the investigation of this question, the
theory of particle condensation in field theory had been a basic framework
\cite{nambu,fritzsch,nilles,fritzsch1,bardeen,coleman,miransky,minkowski,clague,
shifman, fukuda, celenza, roberts, pavel}. In an effort to explore this issue,
an alternative to the standard PQCD, including the presence of quark and gluon
condensates in the free vacuum state generating the Wick expansion, has been
considered in Refs.
\cite{mpla,prd,epjc,epjc1,jhep,epjc2,epjc19,ana,hoyer,hoyer1,hoyer2}. \ The
exploration was in great extent motivated by the consideration about that
massless QCD convey the strongest forces in Nature, as well as the free theory
is a massless highly degenerate one for both quarks and gluons. This fact
rises question about what could be the real intensity of the dimensional
transmutation effect \cite{coleman}. \ The first step was to search for simple
states of the free theory in which large numbers of zero momentum quark and
gluon states were created in the standard free vacuum state, on which, then,
adiabatically connect the interaction \cite{prd,jhep,epjc}. In the start, the
objective was to determine a modification of the standard Feynman rules of QCD
embodying new condensate effects, with the expectation of to attain a
theoretical prediction of superconductivity kind of properties of the particle
mass spectrum underlined in Refs. \cite{fritzsch, fritzsch1}.

\ In Refs. \ \cite{mpla,prd,epjc,epjc1,jhep,epjc2,epjc19,ana}, some
indications about the possible dynamic generation of quark and gluon
condensates were obtained. In particular in Ref. \cite{ana}\ it was
\ restricted the study to consider only the existence of quark condensates,
with the aim of exploring the generation of large values for them. The
motivation was the suspicion about that this mechanism might be able in
generating a variant of the $top$ quark model, as an effective action for
massless QCD. \ In this case, the generating functional $Z$ of the system
introduced in Ref.\cite{epjc19} was transformed to an alternative
representation in which all the effects of the condensates were incorporated
in a new vertex showing two quark and two gluon legs. This representation
implemented the dimensional transmutation effect produced by the modification
done in the free vacuum state. \ The results suggested a technical path that
could evidence a possible strong instability of massless QCD under the
generation of fermion condensates. This outcome could furnish an explanation
of the particle mass hierarchy in a kind of generalized \ Nambu-Jona Lasinio
dynamical symmetry mechanism \cite{nambu,fritzsch, bardeen,miransky}. However,
a drawback of the diagram expansion introduced in Ref. \cite{ana} was that the
new vertex, although being covariant was a nonlocal one. That is, it was not
associated to the four appearing fields defined at the same space-time point.
Moreover, the evaluated two loop corrections to the vacuum energy turned out
to be unbounded from below as a function of the quark condensate parameter, a
fact that shifted the answer of the question about the prediction of dynamical
mass generation to higher loop evaluations. \ \ The undesirable properties of
the new vertex, \ can be traced back to the form of initial state of the free
theory employed to connect the gauge interaction. This wavefunction was a sort
squeezed state constructed with nearly zero momentum quark creations operators
acting on the standard free vacuum. It has a structure similar to the BCS
state in the usual superconductivity theory. The vertex was derived from a
particular representation in which the Feynman gauge was explicitly employed
and the quark free particle states used in the construction were particular
zero momentum states in this special gauge. Therefore, the gauge invariance of
the description was directly broken by the derivation. Henceforth, it came to
the mind the possibility that the connection of the interaction on a specially
defined in a more sophisticated way vacuum state could implement the locality
and gauge invariance of the functional integral action in the ending theory.
In particular coordinate dependent states in this same gauge exist that can be
imagined to allow the mentioned construction.

In this work we start form a the results for the generating functional \ for
the massless QCD in which quark and gluon condensate parameters were
introduced in the free vacuum, for afterwards construct the Wick expansion in
Ref. \cite{epjc19}. \ The resulting scheme included integrals over auxiliary
boson and fermion parameters which appeared in the process of linearly
representing the quadratic forms in the sources introduced by the inclusion of
the gluon and quark condensates. \ Then, it is firstly observed that the
simple promotion of the gluon and quark condensate parameters to be space
dependent functions \ makes the action defining the generating functional a
local and gauge invariant one. \ Further, arguments are presented suggesting
that the connection of the interaction in some sort of more general and yet
undetermined initial modified vacuum state, could show the functional to be
equivalent to massless QCD, after the adiabatic connection of the interaction.
\ Afterwards, the discussion in the work is restricted to the case of
\ retaining only the fermion condensate auxiliary functions. \ The study of
the case of the gluon condensate functions has also appreciable interest in
itself for to explore its possibilities in the description of confinement and
low energy Physics. \ However, the issue of the quark mass generation
possibility is closer to the fermion condensate parameters and we decided to
consider it first here. \ \ In this case, the Gaussian dependence of the quark
auxiliary functions can be integrated out to give a theory action given by the
massless QCD one plus six terms (one for each quark flavour). They implement
the above mentioned local and gauge completions of the non local and gauge non
invariant action obtained in Ref. \cite{ana}. \

The strongest support we had found for the physical relevance of these new
terms simply comes form their structure; it seems that they are allowed
counterterms of the massless QCD Lagrangian. In other words, if not obstructed
by a subtle point not noticed by us, it seems that such a contributions can
appears in the effective action of massless QCD after finite renormalizations
of the four legs 1PI vertex determining the two gluon and two fermion fields
terms in this expansion.\ Curiously, the new terms does not obstacle power
counting renormalizability. This apparently strange property can be understood
by taking into account that the new Lagrangian term also includes a
contribution to the quark free propagators, which acquire a more converging
behaviour $1/p^{2}$ at large momenta. Thus, the theory remains being power
counting renormalizable. The above mentioned properties furnish an intrinsic
interest to the theory resulting after simply adding the new local and
renormalizable terms to the massless QCD action, as an alternative to the
standard QCD.

Further, the Feynman diagrams associated to the new vertices and quark
propagators are defined. As before noticed masses for the free fermions are
defined by the new coupling constants associated to each of them. One curious
outcome is that the larger the coupling (to be called from now on "flavour
condensate" couplings) associated to the action terms for each flavour, the
smaller the quark mass becomes. That is, the top quark shows the smaller
condensate coupling. The modified Feynman rules determined by the new vertices
are defined. The gluon free propagator remains unaltered, but the fermion ones
decompose in two terms showing the same massive pole. One of them resembles a
scalar particle propagator and the other shows a Dirac propagator structure
but including the same massive pole in addition to the usual zero mass one. We
postpone the study of the renormalization (the behaviour of the running strong
and flavour condensate couplings) and start here to explore the first
implications of the analysis by evaluating the gluon self-energy in the
$g^{2}$ order in the gauge coupling and all order in the flavour condensate
ones. The result for the gluon polarization operator became transversal as
implied by the gauge invariance for this second order in the coupling
calculation when all orders of the flavour couplings are included. Further,
the transversal part of the polarization operator is employed to evaluate the
vacuum potential (taking all the mean field values equal to zero) as a
function of the flavour condensate couplings. The potential in this case, at
variance with the one evaluated in \cite{ana} resulted to be bounded from
below. Moreover, for small coupling values, in the here considered
approximation, it shows a dynamical symmetry breaking generating non vanishing
values for the quark masses. Since the renormalization group properties of the
theory are delayed to a next study, it is not possible to make here clear cut
phenomenological predictions. However, as a first exploration, the results for
the potential were employed to select the $g$ coupling and the dimensional
regularization scale parameter $\mu$ in order to fix the mass for the top
quark nearly to 173 $GeV$. That is, the flavour condensate couplings are
suggested as playing a role similar to a Higgs field in the approach. In this
calculation the values of the coupling $g^{2}$ and the scale parameter $\mu$
result to be far from satisfying the known connection between the QCD running
coupling and $\mu$. However, it can be taken into account that in the
considered two loop calculation, all the six quark contributions are
"uncoupled". That is, all propagator entering a diagram have the same flavour.
Therefore, the hierarchy of quark masses, if really implied by the theory, as
we currently suspect, is to be expected only after including higher loop terms
in the potential. In this case mixing among different quark species has the
chance of determining a realistic dynamical symmetry breaking effect. This
view about the need of including higher loop corrections coincides with the
one exposed in Ref. \cite{jhep}, where it was remarked that this requirement
becomes a natural one in a situation in which bound state effects are needed
in determining a breaking of the flavour symmetry. In our considered case, to
predict the quark mass hierarchy, it should occurs that a minimal effective
potential appears for a large value of a single flavour quark condensate,
while the other ones become very much smaller in size, and this minimum lays
below another possible one in which various quarks condensate couplings get
equal values. This situation only could have the chance of happening for the
considered expansion, when the potential includes terms in which different
sorts of quark propagators are present. Therefore, it is possible to conclude
that the low values of the coupling got in the evaluation of the potential
done here, is not yet reflecting a drawback for the analysis to be valid in
describing the mass hierarchy. According to the above remarks, the next
important steps in the search seems to be the study of the renormalization
group properties of the expansion and the evaluation of three loops
corrections to the potential for determining whether or not a quark mass
hierarchy follows.

\ The presentation proceeds as follows. \ In Section II the proposal of the
local and renormalizable gauge theory including gluon and quark condensates
and possibly being equivalent to massless QCD is presented. The case of
retaining only the quark condensate is there shown to \ reduce to a modified
local and renormalizable massless QCD action. Section III is devoted to
evaluate the gluon polarization operator in second order $g^{2}$ of the strong
coupling and all orders in the flavour coupling parameters defining the new
vertices, and to check its transversality. Finally, Section IV making use of
the transversal part of the polarization operator, presents the evaluation of
the effective potential in the two loop approximation and all orders in the
new interaction parameters. In the Summary the results are reviewed and
possible extensions of the work commented.

\section{A local and renormalizable modified QCD}

Let us start by reviewing the main elements of the previous work which will be
employed in the following discussion. The generating functional of massless
QCD written in Euclidean variables as in Ref. \cite{epjc19} \ has the
expression
\begin{align}
Z[j,\eta,\overline{\eta},\xi,\overline{\xi}]  &  =\frac{I[j,\eta
,\overline{\eta},\xi,\overline{\xi}]}{I[0,0,0,0]},\nonumber\\
I[j,\eta,\overline{\eta},\xi,\overline{\xi}]  &  =\exp(V^{int}[\frac{\delta
}{\delta j},\frac{\delta}{\delta\overline{\eta}},\frac{\delta}{-\delta\eta
},\frac{\delta}{\delta\overline{\xi}}\frac{\delta}{-\delta\xi}]){\small \times
}\label{Wick}\\
&  \exp(\int\frac{dk}{(2\pi)^{D}}j(-k)\frac{1}{2}D(k)j(k)){\small \times
}\nonumber\\
&  \exp(\sum_{f}\int\frac{dk}{(2\pi)^{D}}\overline{\eta}_{f}(-k)G_{f}%
(k)\eta_{f}(k)){\small \times}\nonumber\\
&  \exp(\int\frac{dk}{(2\pi)^{D}}\overline{\xi}(-k)G_{gh}(k)\xi(k)){\small ,}%
\nonumber\\
f  &  =1,2,...6.
\end{align}
The functional is associated to the action $S_{g}$ depending on the gauge
interaction coupling $g$. \ $S_{0}$ for $g=0$ defines the free action. $S_{g}$
and the vertex part \ Lagrangian $V^{int}$ are defined in terms of the six
quark $\Psi_{f}$ ($f=1,...,6$), gluon $A$ and ghost $\chi$ fields in the form
\begin{align}
S_{g}  &  =\int dx(-\frac{1}{4}F_{\mu\nu}^{a}F_{\mu\nu}^{a}-\frac{1}{2\alpha
}\partial_{\mu}A_{\mu}^{a}\partial_{\nu}A_{\nu}^{a}+\overline{c}%
^{a}\overleftarrow{\partial}_{\mu}D_{\mu}^{ab}c^{b}-\sum_{f}\overline{\Psi
}_{f}^{i}\text{ }i\gamma_{\mu}D_{\mu}^{ij}\Psi_{f}^{j}),\text{ \ }\\
V^{int}  &  =S_{g}-S_{0}, \label{vint}%
\end{align}
where the field intensity, and covariant derivatives follow the conventions
\begin{align}
F_{\mu\nu}^{a}  &  =\partial_{\mu}A_{\nu}^{a}-\partial_{\nu}A_{\mu}%
^{a}-gf^{abc}A_{\mu}^{b}A_{\nu}^{c},\nonumber\\
D_{\mu}^{ij}  &  =\partial_{\mu}\delta^{ij}+ig\text{ }A_{\mu}^{a}T_{a}%
^{ij},\ \ \ \ D_{\mu}^{ab}=\partial_{\mu}\delta^{ab}+gf^{abc}\text{ }A_{\mu
}^{c},\\
\{\gamma_{\mu},\gamma_{\nu}\}  &  =-2\delta_{\mu\nu},\ \ \ \ [T_{a}%
T_{b}]=if^{abc}T_{c}.\nonumber
\end{align}
The gluon and quark condensate parameters representing the modified vacuum
free state used to construct the Wick expansion enter the generating
functional only though the free quark and gluon propagators, which have the
forms
\begin{align}
D_{\mu\nu}^{ab}(k)  &  =\delta^{ab}(\frac{1}{k^{2}}(\delta_{\mu\nu}%
-(1-\alpha)\frac{k_{\mu}k_{\nu}}{k^{2}})\theta_{N}(k)+C_{g}\delta^{D}%
(k)\delta_{\mu\nu}),\nonumber\\
G_{f}^{ij}(k)  &  =\delta^{ij}{\large (}\frac{\theta_{N}(k)}{m+\gamma_{\mu
}k_{\mu}}+C_{f}\delta^{D}(k)I{\large )},\nonumber\\
G_{gh}^{ab}(k)  &  =\delta^{ab}\frac{\theta_{N}(k)}{k^{2}}.
\label{propagators}%
\end{align}
\ The derivation considered the possibility of quarks already having the mass
$m$ appearing. \ These \ expressions are solutions of the tree approximation
for the propagators, since the 4-dimensional Delta functions is a solution of
the homogeneous Dirac and Klein-Gordon equations, because at zero momentum
both equations vanish. As discussed in the \ Ref. \cite{epjc1}\ , the
functions $\theta_{N}(k)$ regularize all the propagators at zero value of the
momentum as required by the gauge theory quantization \cite{nakanishi}.\ \ The
notation of Ref. \cite{muta} is closely followed along the work. The Einstein
summation notation in the flavour indices is employed and a flavour (or other
kind of index) between parenthesis means that it does not follow the Einstein
convention. \

In Refs. \cite{mpla,prd,epjc,epjc1,jhep,epjc2,epjc19,ana}, the alternative
form of the functional integral (\ref{Wick}) for massless QCD was obtained
after modifying the free vacuum, by creating in it gluons, quark and ghosts of
nearly zero momentum. These changes of the initial condition in the process of
connecting the interaction, after using the rules for constructing the Wick
expansion (\cite{gasioro}), \ led to modified free propagators in
(\ref{propagators}). Some quantities had been evaluated with these modified
expansion (\cite{epjc1,jhep,epjc2}) and its general gauge invariance was
argued \cite{epjc1,jhep,epjc2}. It was possible to check that if the gluon
condensate parameter defined here is chosen to match the estimated value of
the more usual gluon condensation parameter $<g^{2}G^{2}>,$ then the masses of
the light quarks are predicted to be the constituent quark mass of nearly
$333$ MeV. In addition, sample loop calculations of the vacuum energy as a
function of the condensate parameters were done in Refs.
\cite{epjc1,jhep,epjc2}, in one loop order or by employing the ladder
approximations for the quark and gluon propagators. They gave indications of a
dynamical generation of the quark and gluon condensates. For gluons the one
loop result gave a form of the potential similar to the potential in the early
chromomagnetic Savvidi models of confinement.

However, the working expression (\ref{Wick}) of the generating functional
shows two limitations, one of theoretical nature and another technical one.
The technical one is associated with the fact that in the new contributions
existed singularities requiring of \ a special regularization process to be
well defined \cite{epjc1}. \ The theoretical one can be described as follows.
In Ref. \cite{epjc19} we linearized the quadratic terms in the gluon and quark
sources by means of introducing Gaussian integration over auxiliary fields.
This led to a functional integral in which interesting dimensionally
transmuted gluon and quark propagators already appeared. Next, in Ref.
\cite{ana} we restricted the discussion to the case in which only the quark
condensate parameter were retained, seeking to transform the modified Feynman
expansion in a more helpful form by integrating over the auxiliary parameters.
It was possible to transform the generating functional by mapping the effect
of the condensate, from incorporating a Delta function term in the free
propagator, to the addition of a new term in the massless QCD action
associated to a vertex, showing two gluon and two quark legs. However, the
form of this vertex, although being Lorentz and global color invariant was not
local in space-time, since it had the form of a double integral over the
product of two pairs of fields evaluated at two different space-time points.
This effect was a direct consequence of the particular Dirac Delta structure
in the momentum space of the free propagator.

In this section we attempt to modified the construction of the modified
expansion in order to overcome both of the problems. The starting idea for
this purpose was suggested by the form of the generating functional obtained
in Refs. \cite{epjc19,ana}. \ In beginning, let us consider the result for $Z$
\ given in Ref. \cite{epjc19} incorporating gluon as well as quark zero
momentum condensates in the free ground state before connecting the
interaction. The expression for $Z$ is given by Eq. (12) of Ref. \cite{epjc19}
in the form
\begin{align}
Z[j,\eta,\overline{\eta},\xi,\overline{\xi}]  &  =\frac{I[j,\eta
,\overline{\eta},\xi,\overline{\xi}]}{I[0,0,0,0]},\nonumber\\
I[j,\eta_{f},\overline{\eta}_{f},\xi,\overline{\xi}]  &  =\frac{1}%
{\mathcal{N}}\int\int d\alpha d\overline{\chi}d\chi\exp{\large [}-\sum
_{f}\overline{\chi}_{f,u}^{i}\chi_{f,u}^{i}-\frac{\alpha_{\mu}^{a}\alpha_{\mu
}^{a}}{2}{\large ]}\nonumber\\
&  \exp{\large [}V^{int}{\large [}\frac{\delta}{\delta j}{\small +(}%
\frac{2C_{g}}{(2\pi)^{D}}{\small )}^{\frac{1}{2}}{\small \alpha,}\frac{\delta
}{\delta\overline{\eta}}{\small +(}\frac{C_{f}}{(2\pi)^{D}}{\small )}%
^{\frac{1}{2}}{\small \chi}_{f}{\small ,}\frac{\delta}{-\delta\eta}%
{\small +(}\frac{C_{f}}{(2\pi)^{D}}{\small )}^{\frac{1}{2}}\overline{\chi}%
_{f}{\small ,}\frac{\delta}{\delta\overline{\xi}}\frac{\delta}{-\delta\xi
}{\small ,\alpha,}\overline{\chi}{\small ,\chi{\large ]]}\times}%
\label{shifted}\\
&  \exp{\large [}\int\frac{dk}{(2\pi)^{D}}{\Large (}{\small j(-k)}\frac{1}%
{2}{\small D}^{F}{\small (k)j(k)+}\sum_{f}\overline{\eta}_{f}{\small (-k)G}%
^{F}{\small (k)\eta}_{f}{\small (k)]+}\overline{\xi}{\small (-k)G}_{gh}%
^{F}{\small (k)\xi(k)]{\Large )}{\large ]},}\nonumber\\
\text{ }\chi &  =(\chi_{1},\chi_{2},...\chi_{f}...,\chi_{6}),\text{
\ \ }\overline{\chi}=(\overline{\chi}_{1},\overline{\chi}_{2},...\overline
{\chi}_{f}...,\overline{\chi}_{6}),\\
d\alpha d\overline{\chi}d\chi &  =%
{\displaystyle\prod\limits_{a,\text{ }\mu}}
d\alpha_{\mu}^{a}%
{\displaystyle\prod\limits_{f,\text{ }i,\text{ }r}}
d\overline{\chi}_{f,\text{ }r}^{i}d\chi_{f,\text{ }r}^{i}.
\end{align}
where the vertex interaction Lagrangian $V^{int}$ is defined in (\ref{vint})
and in which the propagators for gluons, quarks and ghost fields are the usual
Feynman ones as indicated by the superscript $F$. \ The whole expression is a
Gaussian mean value over the gluon and quark auxiliary parameters $\alpha,$
$\overline{\chi}$ ,$\chi$. \ They are constants in coordinate space, $\alpha$
has Lorentz and color gluon indices and $\chi$ and $\overline{\chi}$ have
spinor and quark color indices. Since the Delta function terms were linked
with the zero momentum Fourier components of the sources, the number of
auxiliary Gaussian variables required to represent in a linear form these zero
momentum components of the sources is reduced to the small number of the
indices of the internal field components.

Before continuing, let us rewrite the expression \ for the vertex Lagrangian
$V^{int}$ appearing in the above relation in order to discuss the coming
points. \ To simplify the writing let us define a notation absorbing the
condensate coefficients in new defined parameters as follows%

\[
\phi_{\mu}^{a}={\small (}\frac{2C_{g}}{(2\pi)^{D}}{\small )}^{\frac{1}{2}%
}\alpha_{\mu}^{a},\text{ \ }\beta_{f,r}^{i}={\small (}\frac{C_{f}}{(2\pi)^{D}%
}{\small )}^{\frac{1}{2}}\chi_{f,r}^{i},\text{ \ }\overline{\beta}_{f,r}%
^{i}={\small (}\frac{C_{f}}{(2\pi)^{D}}{\small )}^{\frac{1}{2}}\overline{\chi
}_{f,r}^{i}.
\]
Now, let us call $S_{g}^{\ast}$ , $S_{0}^{\ast}$ the actions $S_{g}$ , $S_{0}$
after being evaluated in the shifted fields which define $V_{int}$ in the
formula (\ref{shifted}) through $V_{int}=S_{g}^{\ast}-S_{0}^{\ast}$. \ These
quantities, as expressed in terms of the fields and the above redefined
condensate parameters, have the expressions
\begin{align}
S_{g}^{\ast}  &  =S_{g}{\small [}A{\small +(}\frac{2C_{g}}{(2\pi)^{D}%
}{\small )}^{\frac{1}{2}}{\small \alpha,}\Psi_{f}{\small +(}\frac{C_{f}}%
{(2\pi)^{D}}{\small )}^{\frac{1}{2}}{\small \chi}_{f}{\small ,}\overline{\Psi
}_{f}{\small +(}\frac{C_{f}}{(2\pi)^{D}}{\small )}^{\frac{1}{2}}\overline
{\chi}_{f}{\small ,}\frac{\delta}{\delta\overline{\xi}},\frac{\delta}%
{-\delta\xi}{\small ]}\nonumber\\
&  =\int dx{\Large [-}\frac{1}{4}F_{\mu\nu}^{a}(A+\phi)F_{\mu\nu}^{a}%
(A+\phi)\nonumber\\
&  -\frac{1}{2\alpha}\partial_{\mu}A_{\mu}^{a}\partial_{\nu}A_{\nu}^{a}%
-\sum_{f}\overline{\Psi}_{f}^{i}\text{ }i\gamma_{\mu}D_{\mu}^{ij}\Psi_{f}%
^{j}-\overline{c}^{a}\partial_{\mu}D_{\mu}^{ab}(A+\phi)c^{b}\nonumber\\
&  -\sum_{f}{\Large (}\overline{\Psi}_{f}^{i}\text{ }i\gamma_{\mu}D_{\mu}%
^{ij}(A+\phi)\Psi_{f}^{j}+\overline{\Psi}_{f}^{i}\text{ }i\gamma_{\mu}D_{\mu
}^{ij}(A)\beta_{f}^{j}+\overline{\beta}_{q}^{i}\text{ }i\gamma_{\mu}D_{\mu
}^{ij}(A)\Psi_{q}^{j}{\Large )}\nonumber\\
&  -\sum_{f}{\Large (}\overline{\beta}_{f}^{i}\text{ }i\gamma_{\mu}\text{
}ig\phi_{\mu}^{a}T_{a}^{ij}\beta_{f}^{j}+\nonumber\\
&  \overline{\Psi}_{f}^{i}\text{ }i\gamma_{\mu}\text{ }ig\phi_{\mu}^{a}%
T_{a}^{ij}\beta_{f}^{j}+\overline{\beta}_{f}^{i}\text{ }i\gamma_{\mu}\text{
}ig\text{ }\phi_{\mu}^{a}T_{a}^{ij}\Psi_{f}^{j}{\Large )],}\\
S_{0}^{\ast}  &  =S_{0}{\small [}A{\small +(}\frac{2C_{g}}{(2\pi)^{D}%
}{\small )}^{\frac{1}{2}}{\small \alpha,}\Psi{\small +(}\frac{C_{f}}%
{(2\pi)^{D}}{\small )}^{\frac{1}{2}}{\small \chi}_{f}{\small ,}\overline{\Psi
}{\small +(}\frac{C_{f}}{(2\pi)^{D}}{\small )}^{\frac{1}{2}}\overline{\chi
}_{f}{\small ,}\frac{\delta}{\delta\overline{\xi}},\frac{\delta}{-\delta\xi
}{\small ]}\nonumber\\
&  =\int dx\text{ }{\large [}-\frac{1}{4}(\partial_{\mu}A_{\nu}^{a}%
-\partial_{\nu}A_{\mu}^{a})(\partial_{\mu}A_{\nu}^{a}-\partial_{\nu}A_{\mu
}^{a})-\frac{1}{2\alpha}\partial_{\mu}A_{\mu}^{a}\partial_{\nu}A_{\nu}%
^{a}-\overline{c}^{a}\partial^{2}c^{a}\nonumber\\
&  -\sum_{f}\overline{\Psi^{i}}_{f}\text{ }i\gamma_{\mu}\partial_{\mu}\Psi
_{f}^{i}{\large ]}%
\end{align}
\ \ \ One important point in the above expressions is related with the last
two terms in $S_{g}^{\ast}.$ Note that these are the only \ linear in the
radiation fields terms appearing and that they vanish when the coupling is set
to zero. Since the condensate auxiliary parameters are constants in
space-time, it follows that these terms are just proportional to the zero
momentum \ Fourier components of the entering kind of radiation field.
Therefore, when this terms are expressed in terms of the functional
derivatives of the corresponding field sources, they are proportional to the
functional derivative over the zero momentum component of the sources.
Therefore, their action over the free generating functionals in
\ (\ref{shifted}) vanish because the functionals are not depending of these
components, thanks to the Nakanishi zero momentum regularization appearing
factors in \ (\ref{Wick}). \ Since the condensate auxiliary parameters are
constants they do not appear in $S_{0}^{\ast}$ which due to the masslessness
of the theory is defined by pure differential operators which action on them vanish.

Henceforth, the action to be employed in what follows is%

\begin{align}
S_{g}^{\ast}  &  =S_{g}^{\ast}[A{\small ,}\Psi{\small ,}\overline{\Psi
}{\small ,}c,\overline{c},{\small \alpha,\chi,\overline{\chi}]}\nonumber\\
&  =\int dx\text{ }{\Large [}-\frac{1}{4}F_{\mu\nu}^{a}(A+\phi)F_{\mu\nu}%
^{a}(A+\phi)\nonumber\\
&  -\frac{1}{2\alpha}\partial_{\mu}A_{\mu}^{a}\partial_{\nu}A_{\nu}%
^{a}-\overline{c}^{a}\partial_{\mu}D_{\mu}^{ab}(A+\phi)c^{b})\nonumber\\
&  -\sum_{f}{\large (}\overline{\Psi}_{f}^{i}\text{ }i\gamma_{\mu}D_{\mu}%
^{ij}(A+\phi)\Psi_{f}^{j}+\overline{\Psi}_{f}^{i}\text{ }i\gamma_{\mu}D_{\mu
}^{ij}(A)\beta_{f}^{j}+\overline{\beta}_{f}^{i}\text{ }i\gamma_{\mu}D_{\mu
}^{ij}(A)\Psi_{f}^{j}\nonumber\\
&  +\overline{\beta}_{f}^{i}\text{ }i\gamma_{\mu}\text{ }ig\phi_{\mu}^{a}%
T_{a}^{ij}\beta_{f}^{j}{\large )}{\Large ].}%
\end{align}

These last properties are useful to obtain an alternative functional integral
form for $Z$. \ For this purpose consider the rewriting of this quantity as
follows
\begin{align}
Z[j,\eta,\overline{\eta},\xi,\overline{\xi}] &  =\frac{I[j,\eta,\overline
{\eta},\xi,\overline{\xi}]}{I[0,0,0,0]},\nonumber\\
I[j,\eta,\overline{\eta},\xi,\overline{\xi}] &  =\frac{1}{\mathcal{N}}\int\int
d\alpha\text{ }d\overline{\chi}\text{ }d\chi\exp{\large [}-\sum_{f}%
\overline{\chi}_{f,u}^{i}\chi_{f,u}^{i}-\frac{\alpha_{\mu}^{a}\alpha_{\mu}%
^{a}}{2}{\large ]}\times\nonumber\\
&  \exp{\large [}S_{g}^{\ast}{\small [}\frac{\delta}{\delta j}{\small ,}%
\frac{\delta}{\delta\overline{\eta}}{\small ,}\frac{\delta}{-\delta\eta}%
,\frac{\delta}{\delta\overline{\eta}},\frac{\delta}{-\delta\eta},\alpha
,\chi,\overline{\chi}{\large ]}{\small \times}\nonumber\\
&  \exp{\large [}{\small -}S_{0}{\small [}\frac{\delta}{\delta j}%
{\small ,}\frac{\delta}{\delta\overline{\eta}}{\small ,}\frac{\delta}%
{-\delta\eta}{\small ,}\frac{\delta}{\delta\overline{\xi}}\frac{\delta
}{-\delta\xi}{\large ]}{\small \times}\nonumber\\
&  \exp{\large [}\int\frac{dk}{(2\pi)^{D}}{\Large (}{\small j(-k)}\frac{1}%
{2}{\small D}^{F}{\small (k)j(k)+}\sum_{f}\overline{\eta}_{f}{\small (-k)G}%
^{F}{\small (k)\eta}_{f}{\small (k)+}\overline{\xi}{\small (-k)G}_{gh}%
^{F}{\small (k)\xi(k)]{\Large )]},}%
\end{align}
Then, by returning each of the free generating functions to their functional
integral original expression giving rise to them, it follows the relation
\begin{align}
&  \exp{\small [-}S_{0}{\small [}\frac{\delta}{\delta j}{\small ,}\frac
{\delta}{\delta\overline{\eta}}{\small ,}\frac{\delta}{-\delta\eta}%
{\small ,}\frac{\delta}{\delta\overline{\xi}},\frac{\delta}{-\delta\xi
}{\small ]\times}\int\mathcal{D}[A,\overline{\Psi},\Psi,\overline
{c},c]\nonumber\\
&  \exp{\small [}\int dx{\Large (}\mathcal{L}_{0}{\small [}A{\small ,}%
\Psi{\small ,}\overline{\Psi}{\small ,}c,\overline{c}{\small ]+j(x)A(x)+}%
\sum_{f}({\small \overline{\eta}}_{f}{\small (x)\Psi}_{f}{\small (x)+\overline
{\Psi}}_{f}{\small (x)\eta}_{f}{\small (x)){\Large )}.}\nonumber\\
&  =\int\mathcal{D}[A,\overline{\Psi},\Psi,\overline{c},c]\exp{\large [}\int
dx\text{ }{\Large (}{\small \mathcal{L}_{0}[A,\Psi,\overline{\Psi}%
,c,\overline{c}]-\mathcal{L}_{0}[A,\Psi,\overline{\Psi},c,\overline
{c}]+j(x)A(x)+}\sum_{f}{\large (}{\small \overline{\eta}}_{f}{\small (x)\Psi
}_{f}{\small (x)+\overline{\Psi}}_{f}{\small (x)\eta}_{f}{\small (x){\large )}%
{\Large )]}.}\nonumber\\
&  =\int\mathcal{D}[A,\overline{\Psi},\Psi,\overline{c},c]\exp{\large [}\int
dx\sum_{f}{\large (}{\small \overline{\eta}}_{f}{\small (x)\Psi}%
_{f}{\small (x)+\overline{\Psi}}_{f}{\small (x)\eta}_{f}{\small (x)}%
{\large )].}%
\end{align}

Employing the above expression it follows for $Z$%
\begin{align}
Z[j,\eta,\overline{\eta},\xi,\overline{\xi}] &  =\frac{1}{\mathcal{N}}\int\int
d\alpha d\overline{\chi}d\chi\exp[-\sum_{f}\overline{\chi}_{f,\text{ }r}%
^{i}\chi_{f,\text{ }r}^{i}-\frac{\alpha_{\mu}^{a}\alpha_{\mu}^{a}}%
{2}]\nonumber\\
&  \exp{\large [}{\small S}_{g}^{\ast}{\small [\frac{\delta}{\delta j}%
,\frac{\delta}{\delta\overline{\eta}},\frac{\delta}{-\delta\eta},\frac{\delta
}{\delta\overline{\eta}},\frac{\delta}{-\delta\eta},\alpha,\chi,\overline
{\chi}]{\large ]}\times}\label{Zmod}\\
&  \int\mathcal{D}[A,\overline{\Psi},\Psi,\overline{c},c]\exp{\large [}\int
dx\text{ }{\large (}{\small j(x)A(x)+\sum_{f}{\large (}\overline{\eta}%
_{f}(x)\Psi_{f}(x)+\overline{\Psi}_{f}(x)\eta_{f}(x){\large )}{\Large )}%
}{\large ]}\nonumber\\
&  =\frac{1}{\mathcal{N}}\int\int d\alpha d\overline{\chi}d\chi\exp
{\large [}-\sum_{f}\overline{\chi}_{f,\text{ }r}^{i}\chi_{f,\text{ }r}%
^{i}-\frac{\alpha_{\mu}^{a}\alpha_{\mu}^{a}}{2}{\large ]}\times\nonumber\\
&  \int\mathcal{D}[A,\overline{\Psi},\Psi,\overline{c},c]\exp{\large [}\int
dx\text{ }{\large (}S_{g}^{\ast}{\small [}A{\small ,\Psi,}\overline{\Psi
},c,\overline{c},\alpha,\chi,\overline{\chi}{\small ]+}\nonumber\\
&  {\small j(x)A(x)+\sum_{f}{\large (}\overline{\eta}_{f}(x)\Psi
_{f}(x)+\overline{\Psi}_{f}(x)\eta_{f}(x){\large ))}{\Large {\large ]}.}%
}\nonumber
\end{align}
where the action $S_{g}^{\ast}$ takes the form%
\begin{align}
S_{g}^{\ast} &  =S_{g}^{\ast}[A{\small ,}\Psi{\small ,}\overline{\Psi
}{\small ,}c,\overline{c},{\small \alpha,\chi,\overline{\chi}]}\nonumber\\
&  =\int dx{\large [}{\Large -}\frac{1}{4}F_{\mu\nu}^{a}(A+{\small (}%
\frac{2C_{g}}{(2\pi)^{D}}{\small )}^{\frac{1}{2}}\alpha_{\mu}^{a})F_{\mu\nu
}^{a}(A+{\small (}\frac{2C_{g}}{(2\pi)^{D}}{\small )}^{\frac{1}{2}}\alpha
_{\mu}^{a})\nonumber\\
&  -\frac{1}{2\alpha}\partial_{\mu}A_{\mu}^{a}\partial_{\nu}A_{\nu}%
^{a}-\overline{c}^{a}\partial_{\mu}D_{\mu}^{ab}(A+{\small (}\frac{2C_{g}%
}{(2\pi)^{D}}{\small )}^{\frac{1}{2}}\alpha_{\mu}^{a})c^{b})\nonumber\\
&  -\sum_{f}{\Large (}\overline{\Psi}_{f}^{i}\text{ }i\gamma_{\mu}D_{\mu}%
^{ij}(A+{\small (}\frac{2C_{g}}{(2\pi)^{D}}{\small )}^{\frac{1}{2}}\alpha
_{\mu}^{a})\Psi_{f}^{j}+\overline{\Psi}_{f}^{i}\text{ }i\gamma_{\mu}D_{\mu
}^{ij}(A)\chi_{f}^{j}{\small (}\frac{C_{f}}{(2\pi)^{D}}{\small )}^{\frac{1}%
{2}}+{\small (}\frac{C_{f}}{(2\pi)^{D}}{\small )}^{\frac{1}{2}}\overline{\chi
}_{f}^{i}\text{ }i\gamma_{\mu}D_{\mu}^{ij}(A)\Psi_{f}^{j}\nonumber\\
&  +{\small (}\frac{C_{f}}{(2\pi)^{D}}{\small )(}\frac{2C_{g}}{(2\pi)^{D}%
}{\small )}^{\frac{1}{2}}\overline{\chi}_{f}^{i}\text{ }i\gamma_{\mu}\text{
}ig\alpha_{\mu}^{a}T_{a}^{ij}\chi_{f}^{j}{\large )]}{\Large .}%
\end{align}
This formula is exactly the same \ that was obtained in Ref. \cite{ana} \ when
the gluon condensate parameter vanishes. This can be seen more clearly after
considering that all the derivatives of the quark auxiliary parameters vanish.
\ This alternative expression suggests that the above defined local and
renormalizable modification of QCD has the chance of being equivalent to
massless QCD. \ \ For seeing this possibility it can be observed that the non
locality and lack of gauge invariance of the theory defined by \ Z, are just
determined by the coordinate independence of the auxiliary variables. \ In
other words if $\overline{\chi}_{f}^{i}$, $\chi_{f}^{j}$ and $\alpha_{\mu}%
^{a}$ are formally considered as arbitrary functions of the coordinates, the
action $S_{g}^{\ast}$ (after substracting from it the gauge fixing and the
ghost terms) becomes gauge invariant under the gauge transformations
\begin{align}
w(x) &  =\exp[-i\text{ }\lambda_{a}(x)T^{a}],\nonumber\\
w_{ad}(x) &  =\exp[-i\text{ }\lambda_{a}(x)T_{ad}^{a}],\text{ \ \ \ }%
(T_{ad}^{a})^{ik}\equiv i\text{ }f\text{ }^{iak},\nonumber\\
A_{\mu}^{\prime}(x) &  =w_{ad}A_{\mu}(x)w_{ad}^{-1}-\frac{i}{g}w_{ad}%
(x)\partial_{\mu}w_{ad}(x)^{-1},\text{ \ \ \ }\nonumber\\
\Psi_{f}^{\prime}(x) &  =w(x)\Psi_{f}(x)w(x)^{-1},\overline{\Psi}_{f}^{\prime
}(x)=w(x)\overline{\Psi}_{f}(x)w(x)^{-1},\nonumber\\
c^{\prime}(x) &  =w_{ad}(x)c(x)w_{ad}(x)^{-1},\overline{c}^{\prime}%
(x)=w_{ad}(x)\overline{c}(x)w_{ad}(x)^{-1}\nonumber\\
\phi_{\mu}^{\prime}(x) &  =w_{ad}(x)\phi_{\mu}(x)w_{ad}(x)^{-1},\nonumber\\
\chi_{f} &  =w(x)\chi_{f}(x)w(x)^{-1},\nonumber\\
\overline{\chi}_{f}^{^{\prime}} &  =w(x)\overline{\chi}_{f}^{i}(x)w(x)^{-1}%
,\label{gauge}%
\end{align}
where the auxiliary functions (before being the parameters) are represented as
elements of the SU(3) algebra. The boson fields like $A_{\mu}(x)=A_{\mu}%
^{a}(x)T_{ad}^{a},$ are expressed in terms of the adjoint representation for
the algebra generators $T_{ad}^{a},$ (the subscript $ad$ indicates "adjoint")
and the fermion quantities  in the fundamental one in terms of the Gell-Mann
matrices. \

However, the constant in the coordinates character of the condensate
parameters, followed as a direct consequence of the particular nature in which
the modified free vacuum state including condensed zero momentum gluons and
quarks was constructed. \ Thus, it is possible to conceive a wide range of
alternative ways of defining generalized modified free vacuum wavefunctions
within the manifold of states annihilated by the BRST charge of the free
problem. \ Such a possibility, leads to the expectation about that a procedure
\ can be developed in which the ending outcome can be of the similar form as
(\ref{Zmod}), but in which the \ Gaussian integrations over the condensate
parameters \ become substituted by Gaussian functional integrations over
condensate functions depending of the coordinates. \ \ We had attempted to
determine such an alternative procedure, which validity will shows that the
modified theory is in fact equivalent to massless QCD, but up to now we had
not been able to find it. \ \ \ In this work however, we simply will propose
the generalized expansion and consider the first implications of the local
gauge invariant alternative forms of QCD that they represent.

Therefore, we propose the following expression for the generating functional
of a local and renormalizable gauge theory \ which, as it was argued above,
has the chance of reflecting nonperturbative properties of massless QCD%
\begin{align}
Z[j,\eta,\overline{\eta},\xi,\overline{\xi}] &  =\frac{1}{\mathcal{N}}\int%
\int\mathcal{D}[\alpha,\overline{\chi},\chi]\exp[-\sum_{f}\overline{\chi
}_{f,\text{ }r}^{i}(x)\chi_{f,\text{ }r}^{i}(x)-\frac{\alpha_{\mu}%
^{a}(x)\alpha_{\mu}^{a}(x)}{2}]\times\nonumber\\
&  \int\mathcal{D}[A,\overline{\Psi},\Psi,\overline{c},c]\exp{\large [}\int
dx\text{ }{\large (}S_{g}^{\ast}{\small [}A{\small ,\Psi,}\overline{\Psi
},c,\overline{c},\alpha,\chi,\overline{\chi}{\small ]+}\nonumber\\
&  {\small j(x)A(x)+}\sum_{f}{\large (}{\small \overline{\eta}}_{f}%
{\small (x)\Psi}_{f}{\small (x)+\overline{\Psi}}_{f}{\small (x)\eta}%
_{f}{\small (x){\large ))]}},
\end{align}
in which the action is defined by%

\begin{align}
S_{g}^{\ast} &  =S_{g}^{\ast}[A{\small ,}\Psi{\small ,}\overline{\Psi
}{\small ,}c,\overline{c},{\small \alpha,\chi,\overline{\chi}]}\nonumber\\
&  =\int dx{\Large [}-\frac{1}{4}F_{\mu\nu}^{a}(A+{\small (}\frac{2C_{g}%
}{(2\pi)^{D}}{\small )}^{\frac{1}{2}}\alpha_{\mu}^{a})F_{\mu\nu}%
^{a}(A+{\small (}\frac{2C_{g}}{(2\pi)^{D}}{\small )}^{\frac{1}{2}}\alpha_{\mu
}^{a})\nonumber\\
&  -\frac{1}{2\alpha}\partial_{\mu}A_{\mu}^{a}\partial_{\nu}A_{\nu}%
^{a}-\overline{c}^{a}\partial_{\mu}D_{\mu}^{ab}(A+{\small (}\frac{2C_{g}%
}{(2\pi)^{D}}{\small )}^{\frac{1}{2}}\alpha_{\mu}^{a})\text{ }c^{b}%
)\nonumber\\
&  -\sum_{f}\overline{\Psi}_{f}^{i}\text{ }i\gamma_{\mu}D_{\mu}^{ij}%
(A+{\small (}\frac{2C_{g}}{(2\pi)^{D}}{\small )}^{\frac{1}{2}}\alpha_{\mu}%
^{a})\text{ }\Psi_{f}^{j}\nonumber\\
&  -\sum_{f}\overline{\Psi}_{f}^{i}\text{ }i\gamma_{\mu}D_{\mu}^{ij}%
(A)\chi_{f}^{j}\text{ }{\small (}\frac{C_{f}}{(2\pi)^{D}}{\small )}^{\frac
{1}{2}}-\sum_{f}{\small (}\frac{C_{f}}{(2\pi)^{D}}{\small )}^{\frac{1}{2}%
}\overline{\chi}_{f}^{i}\text{ }i\gamma_{\mu}D_{\mu}^{ij}(A)\Psi_{f}%
^{j}\nonumber\\
&  -\sum_{f}{\small (}\frac{C_{f}}{(2\pi)^{D}}{\small )(}\frac{2C_{g}}%
{(2\pi)^{D}}{\small )}^{\frac{1}{2}}\overline{\chi}_{f}^{i}\text{ }%
i\gamma_{\mu}\text{ }ig\alpha_{\mu}^{a}T_{a}^{ij}\chi_{f}^{j}{\Large ].}%
\end{align}
In the case of the vanishing quark condensate parameters, we expect that the
expansion can lead to interesting physical consequences in the low energy
region near $1$ GeV, were a similar discussion in the previous non local
expansion gave predictions \ for the constituent quark masses of the light
fermions and reproduced the Savvidy chromomagnetic effective potential form,
as a function of the gluon condensate parameter. \ Such results ca be expected
to also come out from the generalized formulation proposed in the mean field \ approximation.

\ However, in what rest of the paper we \ will explore the case being closer
to the possibility for quark mass generation, by simplifying the discussion
fixing to zero the gluon parameter. \

\section{The pure quark condensate case}

\ When the gluon condensate parameter is set to zero, the $Z$ \ functional
takes the simpler form%
\begin{align}
Z[j,\eta,\overline{\eta},\xi,\overline{\xi}] &  =\frac{1}{\mathcal{N}}\int%
\int\mathcal{D}[\alpha,\overline{\chi},\chi]\exp{\large [}-\sum_{f}%
\overline{\chi}_{f,\text{ }r}^{i}(x)\chi_{f,\text{ }r}^{i}(x)-\frac
{\alpha_{\mu}^{a}(x)\alpha_{\mu}^{a}(x)}{2}{\large ]}\times\nonumber\\
&  \int\mathcal{D}[A,\overline{\Psi},\Psi,\overline{c},c]\exp{\large [}\int
dx\text{ }{\large (}S_{g}^{\ast}{\small [}A{\small ,\Psi,}\overline{\Psi
},c,\overline{c},\chi,\overline{\chi}{\small ]+}\nonumber\\
&  {\small j(x)A(x)+\sum_{f}{\large (}\overline{\eta}_{f}(x)\Psi
_{f}(x)+\overline{\Psi}_{f}(x)\eta_{f}(x){\large ))]}},\label{Zquark}%
\end{align}
where now the action gets the expression%
\begin{align}
S_{g}^{\ast} &  =S_{g}^{\ast}[A{\small ,}\Psi{\small ,}\overline{\Psi
}{\small ,}c,\overline{c},{\small \chi,\overline{\chi}]}\nonumber\\
&  =\int dx\text{ }{\Large [-}\frac{1}{4}F_{\mu\nu}^{a}F_{\mu\nu}^{a}-\frac
{1}{2\alpha}\partial_{\mu}A_{\mu}^{a}\partial_{\nu}A_{\nu}^{a}-\overline
{c}^{a}\partial_{\mu}D_{\mu}^{ab}c^{b})\nonumber\\
&  -\sum_{f}{\Large (}\overline{\Psi}_{f}^{i}\text{ }i\gamma_{\mu}D_{\mu}%
^{ij}\Psi_{f}^{j}+\overline{\Psi}_{f}^{i}\text{ }i\gamma_{\mu}D_{\mu}^{ij}%
\chi_{f}^{j}\text{ }{\small (}\frac{C_{f}}{(2\pi)^{D}}{\small )}^{\frac{1}{2}%
}+{\small (}\frac{C_{f}}{(2\pi)^{D}}{\small )}^{\frac{1}{2}}\overline{\chi
}_{f}^{i}\text{ }i\gamma_{\mu}D_{\mu}^{ij}\Psi_{f}^{j}{\Large )]},
\end{align}
in which $\chi_{f}^{j}$ and $\overline{\chi}_{f}^{i}$ are now space-time
coordinate dependent functions. \ \ Next, similarly as it was done in Ref.
\ \cite{ana}, we can make use of the fact that the \ dependence of the action
$S_{g}^{\ast}$ is quadratic in the \ auxiliary functions $\chi_{f}$ \ \ and
$\overline{\chi}_{f},$ and thus the Gaussian integration can be explicitly
evaluated by solving the Euler equations
\begin{align}
\frac{\delta S_{g}^{\ast}[A,\overline{\Psi},\Psi,\overline{c},c,\overline
{\chi},\chi]}{\delta\overline{\chi}_{f}^{i}} &  =-\chi_{f}^{i}-{\small (}%
\frac{C_{q}}{(2\pi)^{D}}{\small )}^{\frac{1}{2}}i\gamma_{\mu}D_{\mu}^{ij}%
\Psi_{f}^{j}=0,\label{L1}\\
\frac{\delta S_{g}^{\ast}[A,\overline{\Psi},\Psi,\overline{c},c,\overline
{\chi},\chi]}{\delta\chi_{f}^{i}} &  =\overline{\chi}_{f}^{i}+\overline
{\Psi^{j}}_{f}\text{ }i\gamma_{\mu}\overleftarrow{D}_{\mu}^{ji}{\small (}%
\frac{C_{q}}{(2\pi)^{D}}{\small )}^{\frac{1}{2}}=0,\label{L2}\\
D_{\mu}^{ji} &  =\delta^{ji}\partial+igA_{\mu}^{a}T_{a}^{ji},\text{
\ \ }\overleftarrow{D}_{\mu}^{ji}=-\delta^{ji}\overleftarrow{\partial
}+igA_{\mu}^{a}T_{a}^{ji},
\end{align}
for the auxiliary functions $\chi_{f}$ and $\overline{\chi}_{f}$ and
substituting in (\ref{Zquark}). After doing this, the \ generating functional
can be written as%
\begin{align}
Z &  =\frac{1}{\mathcal{N}}\int\mathcal{D}[A,\overline{\Psi},\Psi,\overline
{c},c,]\exp[{S[A,\overline{\Psi},\Psi,\overline{c},c]}],\label{Z01}\\
{S[A,\overline{\Psi},\Psi,\overline{c},c]} &  ={S}_{mqcd}{[A,\overline{\Psi
},\Psi,\overline{c},c]+\text{ }S^{q}[A,\overline{\Psi},\Psi]}\label{ac1}\\
S_{mqcd}[A,\overline{\Psi},\Psi,\overline{c},c] &  =\int dx(-\frac{1}{4}%
F_{\mu\nu}^{a}F_{\mu\nu}^{a}-\frac{1}{2\alpha}\partial_{\mu}A_{\mu}%
^{a}\partial_{\nu}A_{\nu}^{a}-\sum_{f}\overline{\Psi}_{f}^{i}\text{ }%
i\gamma_{\mu}D_{\mu}^{ij}\Psi_{f}^{j}-\overline{c}^{a}\partial_{\mu}D_{\mu
}^{ab}c^{b}),\label{ac2}%
\end{align}
where the action $S_{mqcd}$ is the usual one for massless QCD and as in
Ref.\cite{ana}, new action terms appear, one for each quark flavour $f$.
\ They have the expressions%
\begin{equation}
S^{q}[A,\overline{\Psi},\Psi]=-\sum_{f\text{ }f^{\prime}}\frac{C_{f\text{
}f^{\prime}}}{(2\pi)^{D}}\int dx\overline{\Psi}_{f}^{j}\text{ }i\gamma_{\mu
}\overleftarrow{D}_{\mu}^{ji}\text{ }i\gamma_{\nu}D_{\nu}^{ik}\Psi_{f^{\prime
}}^{k}.\label{ac3}%
\end{equation}

\ Note that we have slightly generalized the writing of the new action by a
noting the gauge invariance and locality is not lost if we consider that the
two quark fields appearing in the terms can have different flavours.
\ \ \ However, if the matrix $C_{f\text{ }f^{\prime}}$ is assumed to be
hermitian, it can always be diagonalized by a unitary transformation in
flavour space. \ \ Thus, we will assume that the matrix is \ diagonal.
Therefore, it follows that in the case in which only the quark parameters are
present, the resulting action \ is a modification of the massless QCD, in
which new local and gauge invariant interaction terms appear, one for each
quark flavour. \ The resulting theory has few properties that are underlined below:

1) The new terms produce four legs vertices \ which curiously do not affect
the renormalizability of the perturbative expansion, \ as it might be expected
due to the dimensional character of the new condensate coupling $\frac{C_{f}%
}{(2\pi)^{D}}$ parameter. This property can be understood after noticing that
the quadratic in the fields parts of the terms contribute to the new free
quark propagator and made it to decrease with the square of the momenta in the
ultraviolet region. \ Thus the power counting renormalizability \ of massless
QCD is retained.

2) \ In addition, these action terms seems to be possible counterterms
appearing in the context of the renormalization of the standard massless QCD,
possibly representing dimensional transmutation effects. Thus, the interesting
possibility arises that a special renormalization procedure implementing a
dimensional transmutation can also validate the theory proposed, as an
effective action for massless QCD. \ The investigation of this question will
be considered elsewhere. \

\ In what follows we will consider the evaluation of the one loop gluon
self-energy and the vacuum energy as a function of the condensate parameters. \

\section{The Feynman expansion}

Let us describe in this section the Feynman expansion for the case in which
only the quark condensate parameters are non vanishing. \ For bookkeeping
purposes we will consider this expansion in \ Minkowski space. Thus, let us
define an action in terms of the fields defined in Minkowski space,
incorporating all the terms appearing in the expressions (\ref{ac1}%
,\ref{ac2},\ref{ac3}) for the Euclidean action obtained in the previous
section. Then, the Minkowski action to be considered in what follows is chosen
in the form as
\begin{align}
S  &  ={\Large [}\int dx(-\frac{1}{4}F_{\mu\nu}^{a}F^{a\mu\nu}-\frac
{1}{2\alpha}\partial_{\mu}A^{a\mu}\partial_{\nu}A^{a\nu}+\overline{c}%
^{a}\partial_{\mu}D^{ab\mu}c^{b}+\sum_{f}\overline{\Psi}_{f}^{i}\text{
}i\gamma^{\mu}D_{\mu}^{ij}\Psi_{f}^{j})\nonumber\\
&  -\sum_{f}\frac{C_{f}}{(2\pi)^{D}}\int dx\overline{\Psi}_{f}^{j}\text{
}\gamma_{\mu}\overleftarrow{D}^{ji\mu}\text{ }\gamma_{\nu}D^{ik\nu}\Psi
_{f}^{k}{\Large ]},\text{ \ }%
\end{align}
which follows the same conventions of Ref. \cite{muta}. \ \ The field
intensity, and covariant derivatives are now defined by \
\begin{align}
F_{\mu\nu}^{a}  &  =\partial_{\mu}A_{\nu}^{a}-\partial_{\nu}A_{\mu}%
^{a}-g\text{ }f^{abc}A_{\mu}^{b}A_{\nu}^{c},\nonumber\\
D_{\mu}^{ij}  &  =\partial_{\mu}\delta^{ij}-i\text{ }g\text{ }A_{\mu}^{a}%
T_{a}^{ij},\ \ \ \ D_{\mu}^{ab}=\partial_{\mu}\delta^{ab}-g\text{ }%
f^{abc}\text{ }A_{\mu}^{c},\\
\{\gamma_{\mu},\gamma_{\nu}\}  &  =2g_{\mu\nu},\ \ \ \ [T_{a}T_{b}]=i\text{
}f^{abc}T_{c}.\nonumber
\end{align}

The Minkowski action has been written with the same conventions employed in
Ref. \cite{muta}\ and thus all the properties of the vertices and free
propagators associated to the \ massless QCD part of the Lagrangian can be
found in this reference. \ Then, let us only consider the definitions of the
new form of the quark propagators and of the additional four legs vertices.
\ Let us also take $C_{f\text{ }f^{\prime}}^{q}$ as Hermitian. Then, as
mentioned before, by a global unitary transformation between the six quark
fields, it can be brought to a diagonal form $C_{f\text{ }f^{\prime}%
}=C_{f\text{ }}\delta_{f\text{ }f^{\prime}}$\ .

\ \ The expression for the quadratic in \ the quark \ field of flavour $f$
$\ $Lagrangian is \
\begin{align}
\mathcal{L}_{0,f}^{F}  &  =\int dx\text{ }\overline{\Psi}_{f}(i\gamma^{\mu
}\partial_{\mu}-\frac{C_{f}}{(2\pi)^{D}}\partial^{2})\Psi_{f}\nonumber\\
&  =-\int dx\text{ }\overline{\Psi}_{f}\text{ }\Lambda_{f}(\partial
)\Psi_{f^{\prime}}^{j}%
\end{align}
Thus, the propagator \ $S_{f}$ $\ $for each of the six quarks is given by the
inverse of the $\Lambda_{f}$ operator appearing above. That is
\begin{equation}
-(i\gamma^{\mu}\partial_{\mu}-\frac{C_{f}}{(2\pi)^{D}}\partial^{2})\text{
}S_{f}(x-y)=\delta(x-y),
\end{equation}
or in terms of their Fourier transform%

\begin{equation}
-(\gamma^{\mu}p_{\mu}+\frac{C_{f}}{(2\pi)^{D}}p^{2})S_{f}(p)=I.
\end{equation}

Therefore \ the quark propagator for each quark flavour $f$ \ is given as%
\begin{align}
S_{f}(p)  &  =\frac{1}{-\gamma_{\nu}p^{\nu}-\frac{C_{f}}{(2\pi)^{D}}p^{2}%
}\equiv\frac{(-\gamma_{\nu}p^{\nu}-\frac{\text{ }C_{f}}{(2\pi)^{D}}%
p^{2})^{rr^{\prime}}\delta^{ii^{\prime}}}{p^{2}(1-(\frac{\text{ }C_{f}}%
{(2\pi)^{D}})^{2}p^{2})}\nonumber\\
&  =\frac{m_{f}}{(m_{f}^{2}-p^{2})}-\frac{m_{f}^{2}}{(m_{f}^{2}-p^{2})}%
\frac{\gamma_{\nu}p^{\nu}}{p^{2}}=S_{f}^{(s)}(p)+S_{f}^{(f)}(p).
\label{quarkprop}%
\end{align}

It can be observed that the free quark propagators show poles of mass $m_{f}=$
$\frac{(2\pi)^{D}}{C_{f}}$ \ which are inversely proportional to the
\ corresponding condensate parameter $C_{f}$. \ \ In addition, each fermion
propagator decomposes in the sum of a "$scalar$" like term $S_{f}^{(s)}(p)$,
being equal to a massive scalar field propagator times the spinor identity
matrix, and a "$fermion$" like component \ $S_{f}^{(f)}(p)$ having the same
spinor structure as the \ Dirac propagator but also showing an additional pole
at the same mass of the $scalar$ like part. \ The expression for free
generating functional for each quark flavour $f$ \ takes the form%
\begin{equation}
Z_{0}^{F}[\eta,\overline{\eta},f]=\exp{\small [i}\int\frac{dp}{(2\pi)^{D}%
}\overline{\eta}_{f}(p)S_{f}(p)\eta_{f}(p)].
\end{equation}

Let us consider now the vertices associated to the new higher than quadratic
terms in the Lagrangian, after assuming flavour matrix $C$ as not yet
diagonalized
\begin{align}
\mathcal{L}_{0}^{C_{f}}  &  =\sum_{f_{1}f_{2}}\frac{\text{ }g\text{ }%
C_{f_{1}f_{2}}}{(2\pi)^{D}}\int dx\text{ }\overline{\Psi}_{f_{1}}^{j}\text{
}T_{a}^{ji}\text{ }(\gamma_{\mu}A_{a}^{\mu}\gamma_{\nu}\partial^{\nu}%
+\gamma_{\nu}\partial^{\nu}\gamma_{\mu}A_{a}^{\mu})\Psi_{f_{2}}^{i}%
+\nonumber\\
&  +\sum_{f_{1}f_{2}}\frac{g^{2}C_{f_{1}f_{2}\text{ }}}{(2\pi)^{D}}\int
dx\overline{\Psi}_{f_{1}}^{j}\text{ }T_{a}^{ji}\text{ }\gamma_{\mu}A_{a}^{\mu
}T_{a}^{ik}\gamma_{\mu}A_{a}^{\mu}\Psi_{f_{2}}^{k}.\text{ }%
\end{align}

After transforming the integrals to momentum space the expression for the
three and four legs vertices can be written as follows
\begin{align}
V_{(r_{1},i_{1},f_{1})((r_{2},i_{2},f_{2})}^{(3)(\mu,a)}(k_{1},k_{2},k_{3})
&  =g\frac{C_{f_{1}f_{2}}}{(2\pi)^{D}}T_{a}^{i_{1}i_{2}}(-(k_{1\alpha}%
\gamma^{\alpha})^{r_{1}s}(\gamma^{\mu})^{sr_{2}}+(\gamma^{\mu})^{r_{1}%
s}(k_{2\alpha}\gamma^{\alpha})^{sr_{2}}),\label{three}\\
V_{(r_{1},i_{1},f_{1})((r_{2},i_{2},f_{2})}^{(4)\text{ }(\mu,a)(\nu,b)}%
(k_{1},k_{2},k_{3},k_{4})  &  =g^{2}\frac{C_{f_{1}f_{2}}}{(2\pi)^{D}}%
T_{a}^{i_{1}i}T_{b}^{ii_{2}}(\gamma^{\mu})^{r_{1}s}(\gamma^{\nu})^{sr_{2}}.
\label{four}%
\end{align}

\begin{figure}[h]
\begin{center}
\hspace*{-0.4cm} \includegraphics[width=6.5cm]{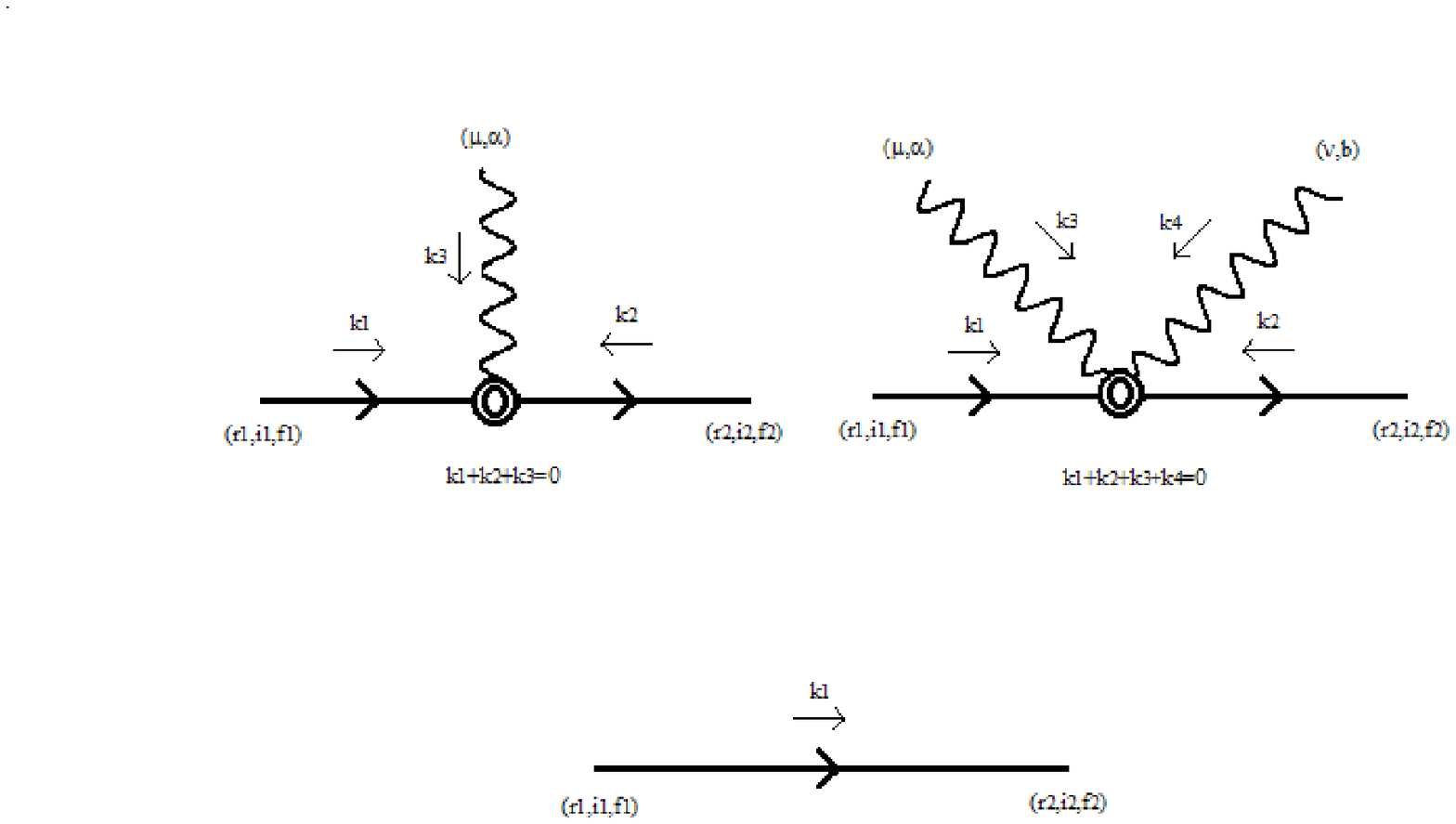}
\end{center}
\caption{ The figure illustrates the diagram associated to the new quark gluon
interaction vertices. The analytic expression of the three legs vertex is
given in (\ref{three}) and the four legs one is defined by (\ref{four}). }%
\label{fig1}%
\end{figure}

The diagrammatic representation of these vertices is illustrates in figure
\ref{fig1}. As usual, the momenta directions coinciding with the direction of
the line associated to a quark propagator means evaluating it in the value of
the momentum. When the direction is opposite the evaluation is in the negative
of the same momentum. As remarked before the rest of the diagrammatic rules
are the same ones as in Ref. \cite{muta}. The quark line, which is the only
one changing its analytic expression in the discussion done here is also depicted.

\section{One loop gluon self-energy evaluation}

\ As it was mentioned before in connection with the extension of the work, the
most important next step seems to consider the renormalization properties of
the proposed modified version of QCD. This issue needs a separate study to be
done. \ However, let us consider here \ two evaluations that can be simply
made finite in the Minimal Substraction scheme (MS), but without adopting yet
definitive renormalization conditions in order to not disregard important
elements not yet clarified. \

\ Specifically, we will evaluate the \ one gluon self energy in the second
order in the color coupling and all orders in the quark condensate parameters
\ $C_{f}=\frac{(2\pi)^{D}}{m_{f}}$ . \ \ The close related two loops
contribution to the vacuum energy in the same approximation, will be also
calculated. In what follows the presented results correspond with calculations
using the Feynman rules of reference \cite{muta} for gluons and ghosts
propagators and all the standard vertices of massless QCD, after complemented
with the before defined rules for the new quark propagator and vertices.
\begin{figure}[h]
\begin{center}
\hspace*{-0.4cm} \includegraphics[width=7.5cm]{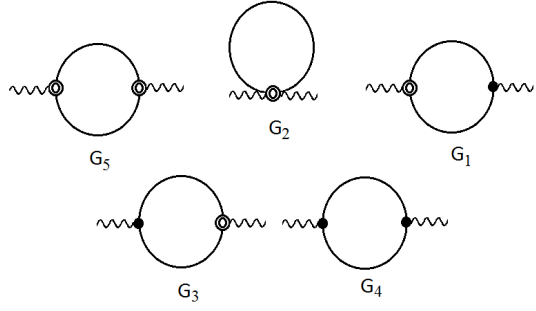}
\end{center}
\caption{ The picture shows the five diagrams associated to the fermion
contribution to the two loop gluon self-energy. They are evaluated in the text
in the same order as they appear from left to right in the figure. }%
\label{fig2}%
\end{figure}

\ The diagrams associated to the mentioned approximation for the self energy
and in which the quarks participate are illustrated in figure $\ref{fig2}$.
\ The terms in which only gluons and ghost propagators and vertices enter are
identical to the ones evaluated in Ref. \cite{muta} are not needed to the
evaluated. In the one loop approximation under consideration the diagrams
entering are sums of \ analytically identical terms for each of the flavours.
Therefore, we will evaluate the contribution to the selfenergy given by only
one flavour. The total contribution is the sum of the same expression obtained
for one quark evaluated in the flavour couplings of each of the quark types.
The expressions for all the appearing vertices and propagators associated to
gluons and ghosts can be found in the exactly the same conventions used here
in Ref. \cite{muta}. \ As an example, the wavy lines design the gluon
propagator as given by.%
\begin{align}
D_{\mu\nu}^{ab}(k)  &  =\frac{\delta^{ab}d_{\mu\nu}(k)}{k^{2}+i\epsilon
},\nonumber\\
d_{\mu\nu}(k)  &  =g_{\mu\nu}-(1-\alpha)\frac{k_{\mu}k_{\nu}}{k^{2}},
\end{align}
and the vertices showing filled dots are the standard quark gluon, ghost gluon
and the two types of gluon vertices as defined in Ref. \ \cite{muta}. The two
kinds of vertices determined by the new terms in the action are represented by
two concentric open circles. \ In order to evaluate the polarization operator
it is convenient to decompose this quantity in its transversal and
longitudinal parts as follows%
\begin{align}
\Pi_{\mu\nu}^{ab}(k)  &  =\Pi_{T}^{ab}(k)(g_{\mu\nu}-\frac{k_{\mu}k_{\nu}%
}{k^{2}})+\Pi_{L}^{ab}(k)\frac{k_{\mu}k_{\nu}}{k^{2}},\\
\Pi_{T}^{ab}(k)  &  =\frac{1}{D-1}(g^{\mu\nu}-\frac{k^{\mu}k^{\nu}}{k^{2}}%
)\Pi_{\mu\nu}^{ab}(k),\\
\Pi_{L}^{ab}(k)  &  =\frac{k^{\mu}k^{\nu}}{k^{2}}\Pi_{\mu\nu}^{ab}(k).
\end{align}

It can be recalled that the fermion propagator (\ref{quarkprop}) is the sum of
one term being proportional to the spinor identity matrix and another one
which is linear in the Dirac matrices. Also the new vertices both have a
quadratic in the Dirac matrices structure. Therefore, it follows that the
expression associated to the diagram \ $G_{5}$ in figure (\ref{fig2}), showing
two of the new three legs vertices, decomposes in the sum of two expressions:
one in which both propagators are of the "$scalar$" type and another in which
both are of the "$fermion$" kind. \ The same happens for the diagram $G_{4}$
in which two usual three legs vertices participate. \ \ The terms associated
to $G_{1}$ in which one of each of the new three legs vertex appears results
in two times the expression in which one propagator is "$scalar$" like \ and
the other one is of the "$fermion$" type.

\subsection{Diagram G$_{5}$}

\subsubsection{Scalar like propagators contribution $\Pi_{5\mu\nu}^{(s)ab}$}

The analytic expression for graph $G_{5}^{(1)},$ \ in which both quark
propagators are of the "$scalar"$ type, after evaluating the color and spinor
traces can be written in the form%
\begin{equation}
\Pi_{5\mu\nu}^{(s)ab}(k)=-8g^{2}\delta^{ab}\int\frac{dp^{D}}{(2\pi)^{D}i}%
\frac{p_{\mu}p_{\nu}-q_{\mu}q_{\mu}+q^{2}g_{\mu\nu}}{(m_{f}^{2}-(p-\frac{q}%
{2})^{2})(m_{f}^{2}-(p+\frac{q}{2})^{2})}.
\end{equation}
which determines the transversal and longitudinal parts as%
\begin{align}
\Pi_{5T}^{(s)ab}(k)  &  =-2\frac{g^{2}\delta^{ab}}{(D-1)}\int\frac{dp^{D}%
}{(2\pi)^{D}i}\frac{4(p^{2}-\frac{(p.q)^{2}}{q^{2}})+(D-1)q^{2}}{(m_{f}%
^{2}-(p-\frac{q}{2})^{2})(m_{f}^{2}-(p+\frac{q}{2})^{2})},\\
\Pi_{5L}^{(s)ab}(k)  &  =-2g^{2}\delta^{ab}\frac{1}{q^{2}}\int\frac{dp^{D}%
}{(2\pi)^{D}i}\frac{4(p.q)^{2}}{(m_{f}^{2}-(p-\frac{q}{2})^{2})(m_{f}%
^{2}-(p+\frac{q}{2})^{2})}.
\end{align}

\subsubsection{Fermion like propagators contribution $\Pi_{5\mu\nu}^{(f)ab}$}

Writing the analytic expression and calculating the traces of the diagram
\ $G_{5}$ in which the quark lines are "$fermion$" like results in the
expression%
\begin{equation}
\Pi_{5\mu\nu}^{(f)ab}(k)=-\frac{1}{2}g^{2}\delta^{ab}m_{f}^{2}\int\frac
{dp^{D}}{(2\pi)^{D}i}\frac{(4p^{2}+q^{2})(4p_{\mu}p_{\nu}-q_{\mu}q_{\nu}%
+q^{2}g_{\mu\nu})-8(p.q)^{2}g_{\mu\nu}}{(p-\frac{q}{2})^{2}(p+\frac{q}{2}%
)^{2}(m_{f}^{2}-(p-\frac{q}{2})^{2})(m_{f}^{2}-(p+\frac{q}{2})^{2})}.
\end{equation}

In this case the integrals for the transversal and longitudinal functions
become%
\begin{align}
\Pi_{5T}^{(f)ab}(k)  &  =-\frac{1}{2}\frac{g^{2}\delta^{ab}}{(D-1)}\int%
\frac{dp^{D}}{(2\pi)^{D}i}\frac{(4p^{2}+q^{2})(4(p^{2}-\frac{(p.q)^{2}}{q^{2}%
})+q^{2}(D-1))-8(p.q)^{2}(D-1)}{(p-\frac{q}{2})^{2}(p+\frac{q}{2})^{2}%
(m_{f}^{2}-(p-\frac{q}{2})^{2})(m_{f}^{2}-(p+\frac{q}{2})^{2})},\\
\Pi_{5L}^{(f)ab}(k)  &  =-\frac{1}{2}g^{2}\delta^{ab}\frac{m_{f}^{2}}{q^{2}%
}\int\frac{dp^{D}}{(2\pi)^{D}i}\frac{4(4p^{2}-q^{2})(p.q)^{2}}{(p-\frac{q}%
{2})^{2}(p+\frac{q}{2})^{2}(m_{f}^{2}-(p-\frac{q}{2})^{2})(m_{f}^{2}%
-(p+\frac{q}{2})^{2})}.
\end{align}

The above integrals which determine the contribution to the diagram G$_{5}$ to
the gluon selfenergy, in general contain more momentum dependent factors in
the denominators that the ones in massless QCD at the same one loop level.
However, those integrals, and the ones appearing in the rest of the
contributions to the self-energy, can be systematically reduced to linear
combinations of one loop scalar integrals, by employing the following
definitions for the factors in the denominator determining the poles of the
integrands
\begin{align}
D_{1}  &  =(p-\frac{q}{2})^{2},\text{ }D_{2}=(p+\frac{q}{2})^{2},\nonumber\\
D_{3}  &  =(m_{f}^{2}-(p-\frac{q}{2})^{2}),\text{ }D_{4}=(m_{f}^{2}%
-(p+\frac{q}{2})^{2}).
\end{align}
These relations can be inverted to express the quantities $p^{2}$ and $p.q$ in
the numerators as linear functions of $q^{2}$ and the $D$ terms in various
ways as follows%

\begin{align}
p^{2}  &  =m_{f}^{2}-\frac{q^{2}}{4}-\frac{1}{2}(D_{3}+D_{4}),\text{
\ \ \ }p.q=\frac{1}{2}(D_{3}-D_{4}),\nonumber\\
p^{2}  &  =\frac{m_{f}^{2}}{2}-\frac{q^{2}}{4}-\frac{1}{2}(D_{4}-D_{1}),\text{
\ \ }p.q=\frac{m_{f}^{2}}{2}-\frac{1}{2}(D_{1}+D_{4}),\nonumber\\
p^{2}  &  =\frac{m_{f}^{2}}{2}-\frac{q^{2}}{4}-\frac{1}{2}(D_{3}-D_{2}),\text{
\ \ \ \ }p.q=-\frac{m_{f}^{2}}{2}+\frac{1}{2}(D_{2}+D_{3}),\nonumber\\
p^{2}  &  =-\frac{q^{2}}{4}+\frac{1}{2}(D_{1}+D_{2}),\text{
\ \ \ \ \ \ \ \ \ \ \ \ }p.q=\frac{1}{2}(D_{2}-D_{1}).
\end{align}

Then, the following identity \ makes the work of decomposing the above Feynman
integrals in simpler ones%
\begin{align}
&  \frac{m_{f}^{4}}{(p-\frac{q}{2})^{2}(p+\frac{q}{2})^{2}(m_{f}^{2}%
-(p-\frac{q}{2})^{2})(m_{f}^{2}-(p+\frac{q}{2})^{2})}\nonumber\\
&  =\frac{m_{f}^{4}}{D_{1}D_{2}D_{3}D_{4}}\nonumber\\
&  =\frac{1}{D_{1}D_{2}}+\frac{1}{D_{3}D_{4}}+\frac{1}{D_{1}D_{4}}+\frac
{1}{D_{2}D_{3}}. \label{decomp}%
\end{align}
Since all \ the integrands in the transversal and longitudinal parts are
functions of \ $p^{2}$ and $p.q$, \ all of them can be expressed as functions
of the $D$ factors and the square of the external momentum $q$, The following
relations between the basic integrals \ resulting in the various evaluations
done here follow%
\begin{align}
0  &  =\int\frac{dp^{D}}{(2\pi)^{D}i}\frac{1}{D_{2}}=\int\frac{dp^{D}}%
{(2\pi)^{D}i}\frac{1}{D_{2}},\nonumber\\
L_{m}(m_{f})  &  =\int\frac{dp^{D}}{(2\pi)^{D}i}\frac{1}{D_{3}}=\int%
\frac{dp^{D}}{(2\pi)^{D}i}\frac{1}{D_{3}},\label{Lm}\\
\int\frac{dp^{D}}{(2\pi)^{D}i}\frac{D_{1}}{D_{3}}  &  =\int\frac{dp^{D}}%
{(2\pi)^{D}i}\frac{D_{2}}{D_{4}}=m_{f}^{2}L_{m}(m_{f}),\nonumber\\
\int\frac{dp^{D}}{(2\pi)^{D}i}\frac{D_{2}}{D_{3}}  &  =\int\frac{dp^{D}}%
{(2\pi)^{D}i}\frac{D_{1}}{D_{4}}=(m_{f}^{2}+q^{2})L_{m}(m_{f}),\nonumber\\
\int\frac{dp^{D}}{(2\pi)^{D}i}\frac{D_{3}}{D_{4}}  &  =\int\frac{dp^{D}}%
{(2\pi)^{D}i}\frac{D_{4}}{D_{3}}=-q^{2}L_{m}(m_{f}),\nonumber\\
\int\frac{dp^{D}}{(2\pi)^{D}i}\frac{D_{3}^{2}}{D_{4}}  &  =\int\frac{dp^{D}%
}{(2\pi)^{D}i}\frac{D_{4}^{2}}{D_{3}}=((q^{2})^{2}+4\frac{m_{f}^{2}q^{2}}%
{D})L_{m}(m_{f}),\nonumber\\
\int\frac{dp^{D}}{(2\pi)^{D}i}\frac{D_{1}^{2}}{D_{4}}  &  =\int\frac{dp^{D}%
}{(2\pi)^{D}i}\frac{D_{2}^{2}}{D_{3}}=((m_{f}^{2}+q^{2})^{2}+4\frac{m_{f}%
^{2}q^{2}}{D})L_{m}(m_{f}).\nonumber
\end{align}

\ Employing the results in references \cite{muta,schroder,smirnov}, performing
the Wick rotation and analytically integrating the appearing Feynman
parametric integrals \ with the use of the Wolfram Mathematica code, \ the
three basic integrals appearing in the above formulae can be evaluated. \ The
tadpole integral \ has the form \cite{muta,schroder}%
\begin{equation}
L_{m}(m_{f})=\int\frac{dp^{D}}{(2\pi)^{D}i}\frac{1}{(m_{f}^{2}+(p-\frac{q}%
{2})^{2})}=\frac{m_{f}^{D-2}}{(4\pi)^{\frac{D}{2}}}\Gamma(1-\frac{D}{2}).
\end{equation}

The massive scalar self-energy \ after making use of the formulae in Ref.
\cite{muta}, performing the Wick rotation and again analytically integrating
the appearing Feynman parametric integral can be written as follows%
\begin{align}
L_{34}(q,m_{f})  &  =\int\frac{dp}{(2\pi)^{D}}\frac{1}{(m_{f}^{2}+(p-\frac
{q}{2})^{2})((m_{f}^{2}+(p+\frac{q}{2})^{2})}\nonumber\\
&  =\frac{1}{(4\pi)^{\frac{D}{2}}}\Gamma(2-\frac{D}{2})m_{f}^{\frac{D}{2}%
-2}\int_{0}^{1}dx(1+x(1-x)\frac{q^{2}}{m_{f}^{2}})^{\frac{D}{2}-2}\nonumber\\
&  =\frac{\Gamma(\epsilon)}{(4\pi)^{2-\epsilon}}(\frac{m_{f}^{2}}{4m_{f}%
^{2}+q^{2}})^{\epsilon}\frac{\sqrt{4m_{f}^{2}+q^{2}}}{q}\times\nonumber\\
&  (B[\frac{1}{2}(1+\frac{q}{\sqrt{4m_{f}^{2}+q^{2}}}),1-\epsilon
,1-\epsilon]-\nonumber\\
&  B[\frac{1}{2}(1-\frac{q}{\sqrt{4m_{f}^{2}+q^{2}}}),1-\epsilon,1-\epsilon]),
\end{align}
where $B[z,a,b]$ \ is the \ Incomplete Gamma Function and as usual
$\epsilon=2-\frac{D}{2}.$%
\begin{equation}
B[z,a,b]=\int_{0}^{z}dtt^{a-1}(1-t)^{b-1}.
\end{equation}

Next, the self-energy term including \ one massive and one massless scalar,
was also evaluated by employing the formula given in Ref. \cite{smirnov} by
after \ the Wick rotation, analytically integrating the appearing parametric
integral. \ \ \ The result becomes
\begin{align}
L_{14}(q,m_{f})  &  =\int\frac{dp}{(2\pi)^{D}i}\frac{1}{(p-\frac{q}{2}%
)^{2}(m_{f}^{2}-(p+\frac{q}{2})^{2})}\nonumber\\
&  =-\frac{\pi^{2-\epsilon}}{(4\pi)^{4-2\epsilon}}\Gamma(\epsilon
)m_{f}^{-2\epsilon}\int_{0}^{1}dx\text{ }x^{-\epsilon}(1-(1-x)\frac{q^{2}%
}{m_{f}^{2}}-i\delta)^{-\epsilon}\nonumber\\
&  =-\frac{\pi^{2-\epsilon}}{(4\pi)^{4-2\epsilon}}\frac{\Gamma(\epsilon
)\Gamma(1-\epsilon)}{\Gamma(2-\epsilon)}(m_{f}^{2}+q^{2})^{-2\epsilon}\text{
}_{2}F_{1}(1-\epsilon,\epsilon,2-\epsilon,-\frac{q^{2}}{m_{f}^{2}-q^{2}}),
\end{align}
in which $_{2}F_{1}(a,b,c,z)$ is the Hypergeometric Function%
\begin{equation}
_{2}F_{1}(a,b,c,z)=\sum_{k=0}^{\infty}\frac{(a)_{k}(b)_{k}}{(c)_{k}}%
\frac{z^{k}}{k!},\text{ }(a)_{k}=\frac{\Gamma(a+k)}{\Gamma(a)}.
\end{equation}

The \ massless scalar self-energy integral results in
\begin{align}
L_{12}(q)  &  =\int\frac{dp}{(2\pi)^{D}i}\frac{1}{(p-\frac{q}{2})^{2}%
(p+\frac{q}{2})^{2}}\nonumber\\
&  =-2^{4\epsilon-5}\pi^{\epsilon-\frac{3}{2}}(q^{2})^{-\epsilon}\frac
{\Gamma(1-\epsilon)\Gamma(\epsilon)}{\Gamma(\frac{3}{2}-\epsilon)}.
\end{align}

Finally, the various contributions to the polarization operator associated to
the diagram G$_{5}$, for each type of quark flavour $f$, can be written as
explicit functions of the quark masses $m_{f}$ (or the quark condensate
parameter $C_{f}$) and the space time dimension $D$ as follows. The amplitudes
of the transversal components result in the form
\begin{align}
\Pi_{5T}^{(s)ab}(q)  &  =-4\frac{g^{2}\delta^{ab}m_{f}^{2}}{(D-1)}\left[
(4m_{f}^{2}+(D-2)q^{2})L_{34}(q,m_{f})-2L_{m}(m_{f})\right]  ,\\
\Pi_{5L}^{(s)ab}(q)  &  =4g^{2}\delta^{ab}m_{f}^{2}L_{m}(m_{f}),
\end{align}%
\begin{align}
\Pi_{5T}^{(f)ab}(q)  &  =T_{12}^{ab}(q)+T_{34}^{ab}(q)+T_{14}^{ab}%
(q)+T_{23}^{ab}(q),\\
T_{12}^{ab}(q)  &  =0,\\
T_{34}^{ab}(q)  &  =-\frac{g^{2}\delta^{ab}}{2(D-1)}\left(  8(\frac{2}%
{D}-5)\text{ }L_{m}(m_{f})+4(4m_{f}^{2}+(D-2)q^{2})L_{34}(q,m_{f})\right)  ,\\
T_{14}^{ab}(q)  &  =T_{23}^{ab}(q)=-\frac{g^{2}\delta^{ab}}{2(D-1)}%
{\Large [(}2m_{f}^{2}(3-D)-2\frac{m_{f}^{4}}{q^{2}}+2q^{2}(D-2){\Large )}%
L_{14}(q,m_{f})+\\
&  {\LARGE (}(1{\Large +}\frac{q^{2}}{m_{f}^{2}})(6-2D+2\frac{m_{f}^{2}}%
{q^{2}})+4(D+1)+2\frac{m_{f}^{2}}{q^{2}}+2(D-2)\frac{q^{2}}{m_{f}^{2}%
}\nonumber\\
&  -\frac{2}{q^{2}}(m_{f}^{2}+2(1+\frac{2}{D})q^{2}+\frac{(q^{2})^{2}}%
{m_{f}^{2}}{\LARGE )}L_{m}(m_{f}){\Large ]},\nonumber
\end{align}
where the indices $(1,2),(3,4),(1,4)$ and ($2,3)$ here and below will indicate
the contributions associated to the corresponding four terms in the
decomposition (\ref{decomp}) of the\ factor $\frac{1}{D_{1}D_{2},D_{3}D_{4}}$.

The longitudinal components get the expressions
\begin{align}
\Pi_{5L}^{(f)ab}(k)  &  =U_{12}^{ab}(q)+U_{34}^{ab}(q)+U_{14}^{ab}%
(q)+U_{23}^{ab}(q),\\
U_{12}^{ab}(q)  &  =0,\\
U_{34}^{ab}(q)  &  =g^{2}\delta^{ab}4(1+\frac{2}{D})\text{ }L_{m}(m_{f}),\\
U_{14}^{ab}(q)  &  =U_{23}^{ab}(q)=-2g^{2}\delta^{ab}{\Large [(}\frac
{m_{f}^{2}}{2q^{2}}(m_{f}^{2}-q^{2}{\Large )}L_{14}(q,m_{f})+\\
&  {\LARGE (-}\frac{m_{f}^{2}}{2q^{2}}+(1+\frac{2}{D}){\LARGE )}L_{m}%
(m_{f}){\Large ]}.\nonumber
\end{align}

\subsection{Diagram G$_{2}$ \ \ \ }

\ This is the simplest of the calculations. After writing the analytic
expression of the graph by following the Feynman rules and evaluating the
color and spinor traces, \ the polarization operator contribution becomes
\begin{align}
\Pi_{2\mu\nu}^{ab}(q)  &  =-4g^{2}\delta^{ab}g^{\mu\nu}\int\frac{dp^{D}}%
{(2\pi)^{D}i}\frac{1}{m_{f}^{2}-p^{2}}\nonumber\\
&  =-4g^{2}\delta^{ab}L_{m}(m_{f})g^{\mu\nu},\\
\Pi_{2T}^{ab}(q)  &  =\Pi_{2L}^{ab}(k)=-4g^{2}\delta^{ab}L_{m}(m_{f}),
\end{align}
in which the transversal and longitudinal parts becomes equal and proportional
to the scalar massive tadpole integral.

\subsection{Diagram G$_{1}$ \ \ \ }

\ \ In this case, as noted before, since there is one new three legs vertex in
the diagram, in which a product of two Dirac gamma matrices enter and another
of the usual massless QCD which only contains one, the calculation reduces to
two times the one in which one fermion like propagator and one scalar of the
scalar type are employed in the internal lines. \ Following the same steps as
before, the integral giving the self-energy contribution associated to the
quark flavour $f$ takes the form
\begin{equation}
\Pi_{1\mu\nu}^{ab}(k)=2g^{2}\delta^{ab}m_{f}^{2}\int\frac{dp^{D}}{(2\pi)^{D}%
i}\frac{(4p_{\mu}p_{\nu}-q_{\mu}q_{\mu}+(q^{2}+2p.q)g_{\mu\nu})}{(p+\frac
{q}{2})^{2}(m_{f}^{2}-(p-\frac{q}{2})^{2})(m_{f}^{2}-(p+\frac{q}{2})^{2})}.
\end{equation}
After getting from it the transversal part \ and expressing the integrand in
terms of the $D$ functions and $q^{2}$, allow to obtain the analytic
expression of this transversal part in terms of the above given scalar
integrals in the form written below
\begin{align}
\Pi_{1T}^{ab}(q)  &  =(W_{12}^{ab}(q)+W_{34}^{ab}(q)+W_{14}^{ab}%
(q)+W_{23}^{ab}(q)),\\
W_{12}^{ab}(q)  &  =0,\\
W_{34}^{ab}(q)  &  =\frac{g^{2}\delta^{ab}}{(D-1)}\left(  (D-2)\text{ }%
q^{2}+4m_{f}^{2})L_{34}(q,m_{f})+(\frac{4}{D}-6)\text{ }L_{m}(m_{f})\right)
,\\
W_{14}^{ab}(q)  &  =\frac{8g^{2}\delta^{ab}}{D}L_{m}(m_{f}),\\
W_{23}^{ab}(q)  &  =\frac{2g^{2}\delta^{ab}}{(D-1)}{\Large [(}m_{f}%
^{2}(3-D)+q^{2}(D-2)-\frac{m_{f}^{4}}{q^{2}}{\Large )}L_{14}(q,m_{f})+\\
&  \frac{(Dq^{2}+m_{f}^{2})}{q^{2}}L_{m}(m_{f}){\Large ].}\nonumber
\end{align}

Following the same steps as before, the evaluated expression for the
coefficient of the longitudinal component becomes\
\begin{align}
\Pi_{1L}^{ab}(q)  &  =-(X_{12}^{ab}(q)+X_{34}^{ab}(q)+X_{14}^{ab}%
(q)+X_{23}^{ab}(q)),\\
X_{12}^{ab}(q)  &  =0,\\
X_{34}^{ab}(q)  &  =4g^{2}\delta^{ab}(1+\frac{2}{D})\text{ }L_{m}(m_{f}),\\
X_{14}^{ab}(q)  &  =-\frac{8g^{2}\delta^{ab}}{D}L_{m}(m_{f}),\\
X_{23}^{ab}(q)  &  =-2g^{2}\delta^{ab}{\Large [}\frac{m_{f}^{2}}{q^{2}%
}{\Large (}m_{f}^{2}-q^{2})L_{14}(q,m_{f})+(2-\frac{m_{f}^{2}}{q^{2}}%
)L_{m}(m_{f}){\Large ].}%
\end{align}

\ The contribution of the diagram $G_{3}$ exactly coincides with the just
evaluated for $G_{1}.$ This can be seen by performing a mood change of the

momentum integration variable in the expression for G$_{1}$.

\subsection{Diagram G$_{4}$}

\ \ \ The process of evaluation of the diagram $G_{4}$ follows the same steps
as the ones for $G_{5}.$ \ The form of the results for the integral after the
traces are calculated is
\begin{align}
\Pi_{4\mu\nu}^{ab}(k)  &  =\Pi_{1\mu\nu}^{(A)ab}(k)+\Pi_{1\mu\nu}%
^{(B)ab}(k),\\
\Pi_{4\mu\nu}^{(A)ab}(k)  &  =-2g^{2}\delta^{ab}m_{f}^{2}g_{\mu\nu}\int%
\frac{dp^{D}}{(2\pi)^{D}i}\frac{1}{(m_{f}^{2}-(p-\frac{q}{2})^{2})(m_{f}%
^{2}-(p+\frac{q}{2})^{2})}\\
&  =-2g^{2}\delta^{ab}m_{f}^{2}L_{34}(q,m_{f})g_{\mu\nu},\nonumber\\
\Pi_{4\mu\nu}^{(B)ab}(k)  &  =-g^{2}\frac{\delta^{ab}m_{f}^{4}}{2}\int%
\frac{dp^{D}}{(2\pi)^{D}i}\frac{8p_{\mu}p_{\nu}-2q_{\mu}q_{\mu}+(q^{2}%
-4p^{2})g_{\mu\nu}}{(p-\frac{q}{2})^{2}(p+\frac{q}{2})^{2}(m_{f}^{2}%
-(p-\frac{q}{2})^{2})(m_{f}^{2}-(p+\frac{q}{2})^{2})},
\end{align}
and the explicit formulae for the coefficients of the transversal and
longitudinal parts of $\Pi_{4\mu\nu}^{(A)ab}$ are
\begin{equation}
\Pi_{4T}^{(A)ab}(k)=\Pi_{4L}^{(A)ab}(k)=-2g^{2}\delta^{ab}m_{f}^{2}%
L_{34}(q,m_{f}).
\end{equation}

Similarly, the result for the transversal part of the contribution $\Pi
_{1\mu\nu}^{(B)ab}$ in terms of the basic integrals become
\begin{align}
\Pi_{4T}^{ab}(q)  &  =B_{12}^{ab}(q)+B_{34}^{ab}(q)+B_{14}^{ab}(q)+B_{23}%
^{ab}(q),\\
B_{12}^{ab}(q)  &  =-\frac{g^{2}(D-2)}{(D-1)}\delta^{ab}q^{2}L_{12}%
(q,m_{f}),\\
B_{34}^{ab}(q)  &  =-\frac{g^{2}\delta^{ab}}{(D-1)}\left(  2(D-2)\text{ }%
L_{m}(m_{f})+(2m_{f}^{2}(3-D)+2q^{2}(D-2))L_{34}(q,m_{f})\right)  ,\\
B_{14}^{ab}(q)  &  =B_{23}^{ab}(q)=-\frac{g^{2}\delta^{ab}}{2(D-1)}%
{\Large [(}2(2-D)+2\frac{m_{f}^{2}}{q^{2}}{\Large )}L_{m}(m_{f})+{\Large (}%
m_{f}^{2}(3-D)+2q^{2}(D-2)-\frac{2m_{f}^{4}}{q^{2}}{\Large )}L_{14}%
(q,m_{f}){\Large ].}%
\end{align}

Finally, the longitudinal term get the expression%
\begin{align}
\Pi_{4L}^{ab}(q)  &  =C_{12}^{ab}(q)+C_{34}^{ab}(q)+C_{14}^{ab}(q)+C_{23}%
^{ab}(q),\\
C_{12}^{ab}(q)  &  =0,\\
C_{34}^{ab}(q)  &  =2g^{2}\delta^{ab}m_{f}^{2}\text{ }L_{34}(q,m_{f}),\\
C_{14}^{ab}(q)  &  =C_{23}^{ab}(q)=-g^{2}\delta^{ab}m_{f}^{2}{\Large (}%
\frac{m_{f}^{2}}{q^{2}}-1)L_{14}(q,m_{f})+g^{2}\delta^{ab}\frac{m_{f}^{2}%
}{q^{2}}L_{m}(m_{f}){\Large ].}%
\end{align}

\subsection{Gauge invariance and total transversal polarization operator}

After adding all the above evaluated contributions to the longitudinal
component of the selfenergy it follows
\begin{align}
\Pi_{L}^{ab}(q)  &  =\Pi_{5L}^{ab}(q)+\Pi_{2L}^{ab}(q)+\Pi_{1L}^{ab}%
(q)+\Pi_{3L}^{ab}(q)+\Pi_{4L}^{ab}(q)\nonumber\\
&  =0.
\end{align}
Therefore, the transversality property is satisfied by the self-energy as it
should be since it is a Ward identity which must be satisfied by any
correction to the self-energy given by a well defined order in the
perturbation theory.\ \ Since, here we evaluated all the terms in the
expansion which are of order two in the strong coupling being and exact (to
all orders ) \ in the flavour condensate parameters, the transversality should
be satisfied.

\ Further, after also adding all the \ transversal contributions and
simplifying the result,\ the total unrenormalized one loop gluon self-energy
can be written in the form \
\begin{align}
\Pi_{\mu\nu}^{ab}(q)  &  =\Pi_{T}^{ab}(q)(g_{\mu\nu}-\frac{q_{\mu}q_{\nu}%
}{q^{2}}),\\
\Pi_{T}^{ab}(q)  &  =\Pi_{5T}^{ab}(q)+\Pi_{2T}^{ab}(q)+\Pi_{1T}^{ab}%
(q)+\Pi_{3T}^{ab}(q)+\Pi_{4T}^{ab}(q)\\
&  =-\frac{g^{2}\delta^{ab}}{D(D-1)}(D(D-26)+8)L_{m}(m_{f})+\frac{g^{2}%
\delta^{ab}(2-D)}{(D-1)}q^{2}L_{12}(q)+\nonumber\\
&  +\frac{g^{2}\delta^{ab}}{2(D-1)}(-2D(D+17)m_{f}^{2}-9D(2+D)q^{2}%
)L_{34}(q,m_{f}).\nonumber
\end{align}

\ Employing the \ Minimal Substraction renormalization scheme (MS), \ the one
loop divergences can be eliminated by deleting the terms showing poles in the
expansion of the above expression in a Laurent series in the parameter
$\epsilon=\frac{4-D}{2}.$ However, the resulting expression is cumbersome and
since we will not analyze it here, is not written. After adding to $\Pi
_{T}^{ab}(q)$ the free part of the gluon equation of motion, the \ following
expressions for the gluon one loop propagator and its inverse follow%
\begin{align}
D_{\mu\nu}^{ab-1}(q)  &  =\delta^{ab}q^{2}(g_{\mu\nu}-(1-\alpha)\frac{q_{\mu
}q_{\nu}}{q^{2}})-\Pi_{T}^{ab}(q)(g_{\mu\nu}-\frac{q_{\mu}q_{\nu}}{q^{2}})\\
&  =(q^{2}-\Pi_{T}^{ab}(q))(g_{\mu\nu}-\frac{q_{\mu}q_{\nu}}{q^{2}}%
)+\alpha\frac{q_{\mu}q_{\nu}}{q^{2}},\nonumber\\
\Pi_{T}^{ab}(q)  &  =\delta^{ab}\Pi_{T}(q),\\
D_{\mu\nu}^{ab}(q)  &  =\frac{\delta^{ab}}{(q^{2}-\Pi_{T}(q))}(g_{\mu\nu
}-\frac{q_{\mu}q_{\nu}}{q^{2}})+\frac{\delta^{ab}}{\alpha}\frac{q_{\mu}q_{\nu
}}{q^{2}}.
\end{align}

\section{ Vacuum energy}

\ Let us consider below the main steps in the evaluation of the one and two
loops corrections to the effective action which will give the negative of the
vacuum energy as a function of the quark masses \ $m_{f}$ in that approximation.\

\subsection{One loop term}

\ The one loop term reduces to the logarithm of the determinant of the new
form of the inverse quark propagator, which is the result of the functional
integral defining free quark generating functional. \ The calculation of this
term, following usual steps in this special case, is\ sketched below
\begin{align}
\Gamma^{(2)}(m_{f})  &  =\frac{1}{i}\sum_{p}\log\det G^{-1}(p)\nonumber\\
&  =-V^{D}\sum_{p}\frac{dp^{D}}{(2\pi)^{D}i}\log\det G(p)\nonumber\\
&  =-V^{D}\sum_{p}\frac{dp^{D}}{(2\pi)^{D}i}\log\det[\frac{m_{f}}{m_{f}%
^{2}-p^{2}}(1-m_{f}\frac{\gamma_{\mu}p^{\mu}}{p^{2}})]\nonumber\\
&  =-V^{D}\sum_{p}\frac{dp^{D}}{(2\pi)^{D}i}\log\det[\frac{m_{f}}{m_{f}%
^{2}-p^{2}}I]-\frac{V^{D}}{2}\sum_{p}\frac{dp^{D}}{(2\pi)^{D}i}\log\det
[\frac{p^{2}-m_{f}^{2}}{p^{2}}I]\nonumber\\
&  =V^{D}\sum_{p}\frac{dp^{D}}{(2\pi)^{D}i}2N\text{ }\log[m_{f}^{2}%
-p^{2}]+V^{D}\sum_{p}\frac{dp^{D}}{(2\pi)^{D}i}2N\text{ }\log[\frac{p^{2}%
}{m^{2}}].
\end{align}

After taking the derivative the expression over $m_{f}^{2}$, it follows%
\begin{align}
\frac{d}{dm_{f}^{2}}\Gamma^{(0)}(m_{f})  &  =V^{D}2N\int\frac{dq^{D}}%
{(2\pi)^{D}i}\frac{1}{m_{f}^{2}-p^{2}}\nonumber\\
&  =V^{D}2N\text{ }L_{m}(m_{f})\nonumber\\
&  =V^{D}2N\text{ }\frac{\Gamma(1-\frac{D}{2})m_{f}^{D-2}}{(4\pi)^{\frac{D}%
{2}}},
\end{align}
which when integrated over $m_{f}^{2}$ in the interval $(0,m_{f}^{2})$ by
assuming $\epsilon$ being in a region in which its real part is negative,
leads to
\begin{equation}
\Gamma^{(0)}(m_{f})=V^{D}2N\text{ }\frac{\Gamma(\epsilon-1)(m_{f}%
^{2})^{2-\epsilon}}{(4\pi)^{\frac{D}{2}}(2-\epsilon)}.
\end{equation}
The finite part of the above expression in the Laurent series expansion can be
evaluated by employing the expression of $V^{D}$ in terms of the
four-dimensional volume and the dimensional regularization scale parameter
$\mu$%
\begin{equation}
V^{D}=\mu^{4-D}V^{4}=\mu^{2\epsilon}V^{4}.
\end{equation}

Then, the finite part the one loop effective action in the MS scheme becomes
\begin{align}
\Gamma_{fin}^{(0)}(m_{f})  &  =2N\frac{m_{f}^{4}}{(4\pi)^{2}}\log(\frac{m_{f}%
}{\mu})-\frac{m_{f}^{4}}{(4\pi)^{2}}(3-\gamma+\log(4\pi))\nonumber\\
&  =-V_{fin}^{(0)}(m_{f}).
\end{align}
The result after changed its sign, gives the dependence on $m_{f}$ of the
vacuum energy in this approximation. The potential decreases at large mass
values indicating the dynamical generation of the mass in this approximation.
This outcome for the first correction was also obtained in Refs.
\cite{epjc2,epjc19,ana},

\subsection{Two loop terms}

The two loop vacuum energy can be readily evaluated starting from the already
known transversal part. \ This follows, thanks to the formula
\begin{align}
\Gamma^{(2)}(m_{f})  &  =\frac{(D-1)}{2}\int\frac{dq^{D}}{(2\pi)^{D}i}%
\frac{\Pi_{T}^{aa}(q)}{q^{2}}\nonumber\\
&  =\frac{(D-1)(N^{2}-1)}{2}\int\frac{dq^{D}}{(2\pi)^{D}i}\frac{\Pi_{T}%
(q)}{q^{2}},
\end{align}
which follows after noting that the diagrams defining the two loop
approximation for the effective action can be organized as the sum of the
diagrams defining the self-energy contracted with the gluon free propagator.

\noindent After substituting in the above formula the evaluated expression for
$\Pi_{T}$ it follows%

\begin{align}
\Gamma^{(2)}(m_{f})  &  =-\frac{g^{2}(N^{2}-1)}{2D}(D(D-26)+8)\int\frac
{dq^{D}}{(2\pi)^{D}i}\frac{L_{m}(m_{f})}{q^{2}}+\frac{g^{2}\delta^{ab}%
(2-D)}{(D-1)}\int\frac{dq^{D}}{(2\pi)^{D}i}L_{12}(q)+\nonumber\\
&  +\frac{g^{2}(N^{2}-1)}{4}\int\frac{dq^{D}}{(2\pi)^{D}i}(-\frac
{2D(D+17)m_{f}^{2}}{q^{2}}+9D(2-D))L_{34}(q,m_{f}).
\end{align}

In the above formula, the first and the second terms vanish in dimensional
regularization. \ \ The third one can be expressed

as follows
\begin{align}
\Gamma^{(2)}(m_{f})  &  =+\frac{g^{2}9(N^{2}-1)D(2-D)V^{D}}{4}\int\frac
{dq^{D}}{(2\pi)^{D}i}\frac{dp^{D}}{(2\pi)^{D}i}\frac{1}{(m^{2}-q^{2}%
)(m^{2}-p^{2})}\nonumber\\
&  +\frac{g^{2}(N^{2}-1)D(D+17)m_{f}^{2}}{2}\int\frac{dq^{D}}{(2\pi)^{D}%
i}\frac{dp^{D}}{(2\pi)^{D}i}\frac{1}{q^{2}(m^{2}-q^{2})(m^{2}-(p-q)^{2})}.
\end{align}

Each of the two integrals appearing in the first term is equal to the
\ $L_{m}$ in (\ref{Lm})$.$ \ The second term is proportional to a two loop
scalar Master Integral \bigskip given in Ref. \cite{schroder}. Then, the
unrenormalized two loops effective action has the form%

\begin{align}
\Gamma^{(2)}(m_{f})  &  =+\frac{9(N^{2}-1)D(2-D)V^{D}}{4}\frac{\Gamma
^{2}(1-\frac{D}{2})}{(4\pi)^{D}}m_{f}^{2D-4}g^{2}V^{D}\nonumber\\
&  -\frac{(N^{2}-1)D(D+17)}{2}\frac{(D-2)\Gamma^{2}(2+\epsilon)\Gamma
^{2}(1-\frac{D}{2})}{2^{2D+1}(D-3)}g^{2}m_{f}^{2D-4}V^{D}. \label{efaction}%
\end{align}

\ Substituting \ the formulae for the coupling $g$ and the $D$ dimensional
volume $V^{D}$according to
\[
g=g_{0}\mu^{D},\text{ \ \ \ }V^{D}=\mu^{2\epsilon}V^{4},
\]
it follows%
\[
g^{2}m_{f}^{2D-4}V^{D}=g_{0}^{2}V^{4}m_{f}^{4}m_{f}^{-4\epsilon}.
\]
Afterwards, expanding the formula (\ref{efaction}) in Laurent series around
$\epsilon=0$, deleting the pole terms and tending to the limit $\epsilon
\rightarrow0$ leads to the finite value of the two loop contribution to the
effective action in the MS scheme
\begin{align}
\Gamma^{(2)}(m_{f})  &  =-273.18\text{ }\frac{g_{0}^{2}}{4\pi}m_{f}^{4}\text{
}+322.47\text{ }\frac{g_{0}^{2}}{4\pi}\text{ }m_{f}^{4}Log(\frac{m_{f}}{\mu
})-\\
&  132.527\text{ }m_{f}^{4}\text{ }\frac{g_{0}^{2}}{4\pi}\text{ }Log^{2}%
(\frac{m_{f}}{\mu}).
\end{align}

The total effective potential is then given by the expression
\begin{align}
V(m_{f})  &  =0.0656145\text{ }m_{f}^{4}\text{ }+273.18\text{ }m_{f}^{4}\text{
}\frac{g_{0}^{2}}{4\pi}-(0.0379954\ +\\
&  322.47\text{ }\frac{g_{0}^{2}}{4\pi})\text{ }m_{f}^{4}\text{ }%
Log(\frac{m_{f}}{\mu})+132.527\text{ }m_{f}^{4}\text{ }\frac{g_{0}^{2}}{4\pi
}\text{ }Log^{2}(\frac{m_{f}}{\mu}).
\end{align}

\begin{figure}[h]
\begin{center}
\hspace*{-0.4cm} \includegraphics[width=7.5cm]{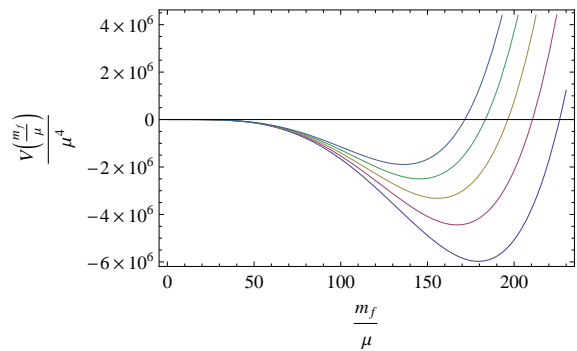}
\end{center}
\caption{ The figure illustrates the effective potential divided by $\mu^{4}$
as function of the ratio $\frac{m_{f}}{\mu}$. At $\mu=1$ GeV various small
values of the coupling constant around $g_{0}=0.0271828$ were chosen to
evidence that the minimum of the potential can be fixed at a $m_{f}$ \ being
close to the top quark mass of $173$ GeV. For higher values of the coupling
the minimum tends to disappear. However, as described in the text, the lack of
the possibility for generating a hierarchy of mass in the considered two loops
approximation, determines the need for improving the evaluation of the
potential, in order to decide whether or not a large and hierarchical mass
generation effect can be predicted at the large values of the strong coupling
observed in Nature. }%
\label{fig3}%
\end{figure}

In Figure \ref{fig3} the potential divided by the fourth power of the
renormalization parameter $\mu$ is plotted as a function of the mass divided
by $\mu$ for various small values of the coupling constant $g_{0}$. For higher
values the coupling minimum of the minimum of the potential tends to
disappear.\ However, as it was remarked before, the two loop approximation is
insufficient to define whether or not the scheme predicts a hierarchical
flavour dynamic symmetry breaking. \ Therefore, the evaluation done should not
be expected to be relevant for answering the main physical question: the
possibility of large quark mass generation at the values of the strong
coupling. \ \ However, \ the results shown in Figure \ref{fig3} indicate that
at small values of the\ gauge coupling, the interaction is able to generate
large masses for the fermions. The figure corresponds to fix as an example
$\mu=1$ $\ $GeV , then it follows that for a \ small coupling near the value
$g_{0}=0.0271828$ the fermion mass gets a value near the top quark mass
$m_{f}=173$ \ GeV. \ \ This \ evaluation suggests the interesting possibility
that, if the theory is in fact equivalent to the massless one, the
\ generation of large mass mechanism could also work for the low coupling
electroweak scale. 

\section{Summary}

We have proposed an improved version of the modified version of QCD discussed
in previous works, which shows local gauge invariance and include the same
kind of gluon and fermion condensate parameters. The analysis done for
constructing the proposal suggests its equivalence with massless QCD, after
considered with a special renormalization procedure designed to implement the
dimensional transmutation effect. In respect to the gluon condensation
properties, the theory should reproduce the previous derivation of the
constituent mass for the light quarks \ given in \cite{epjc,jhep} and the
Savvidy kind of potential as a function of the gluon condensate parameter. The
study of the possible improvement in the predictions determined by the gained
gauge invariance and locality of the description will be considered elsewhere.
For the case of only retaining the fermion condensate parameter, the fermion
auxiliary functions were integrated leading to a theory described by a new
Lagrangian given by the massless QCD one but including a new gauge invariant
term for each quark flavour. The new Lagrangian terms can be considered as
local modifications of the ones derived in Ref. \cite{ana}. In the same form,
they have a similar component constituted by the product of two gluon and two
quark fields, but now evaluated in the same space-time point. \ The new terms
also show components leading to three legs vertices which were absent in the
previous form of the generating functional. The terms determines masses for
all the six quarks which are given by the reciprocal of the new flavour
condensate couplings linked with each quark type. Therefore, the strength of
the condensate couplings decreases with the masses of the associated quarks.
The gluon self-energy was evaluated up to the second order in the gauge
coupling including all orders of the flavour condensate ones. The result,
being associated to a well defined order in the couplings, satisfies the
transversality condition as required by the gauge invariance. In addition, it
is also gauge parameter independent. The transversal part of the self energy
is employed to evaluate the two loop contribution to the effective action at
zero mean fields (minus the vacuum energy) as a function of the flavour
condensate couplings. The transversality and gauge parameter independence of
the gluon self-energy, then also determined the gauge parameter independence
of the evaluated potential. Afterwards, in order to interpret the result, we
assumed that the introduced flavour couplings represent dynamically generated
quantities, as it was effectively the case in the precedent works for the
similar non gauge invariant couplings which motivated their consideration in
this study. Then, in this first approximation, the potential is able to
predict the dynamic generation of quark masses, for sufficiently small values
of the QCD coupling. It can be remarked that the generation happening only for
small couplings at the considered approximation, does not mean a negative
answer to the question about whether or not the scheme can address the quark
mass hierarchy. This is can be understood by noting that in the two loop
order, the diagrams can not yet incorporate lines associated to two different
kinds of flavours. For this appearance to happens, at least three loops
corrections are required. It is reasonable that in order to explain the quark
mass hierarchy as a dynamic flavour symmetry breaking, there should exist
"interference" like effects in the vacuum energy corrections. In them, the
contributions of diagrams showing two or more kinds of fermion lines might
tend to rise the energy of the configurations with equal values of the quark
condensates, making them more energetic that the ones in which one of the
quark condensate parameters gets a large value and the others hierarchical
lower ones. At the moment we have the impression that the considered framework
is ideal to realize the Democratic Symmetry Breaking properties of the mass
hierarchy remarked in Refs. \cite{fritzsch, fritzsch1}. The predictions of the
general discussion including gluon condensates for the low energy processes,
as well as the renormalization properties and the evaluation of three loop
contributions in the case of only having the fermion condensate, are expected
to be considered elsewhere.

\begin{acknowledgments}
I would like to acknowledge N.G. Cabo-Bizet and A. Cabo-Bizet, for their
participation in the conception of the ideas in the work and many helpful
discussions, during and after the finishing of the close connected paper in
Ref. \cite{ana}. I am also indebted by the helpful support received from
various institutions: the Caribbean Network on Quantum Mechanics, Particles
and Fields (Net-35) of the ICTP Office of External Activities (OEA), the
"Proyecto Nacional de Ciencias B\'{a}sicas" (PNCB) of CITMA, Cuba, the
Coordenac\~ao de Aperfeicoamento de Pessoal de N\'ivel Superior (CAPES) of
Brazil and the Postgraduation Programme in Physics (PPGF) of the Federal
University of Par\'a at Bel\'em, Par\'a (Brazil), in which this work was
finished, in the context of a CAPES External Professor Fellowship.
\end{acknowledgments}


\end{document}